\documentclass[10pt,letterpaper]{article}
\usepackage[top=0.85in,left=2.75in,footskip=0.75in]{geometry}

\usepackage{amsmath,amssymb}

\usepackage{changepage}

\usepackage{textcomp,marvosym}

\usepackage{cite}

\usepackage{nameref,hyperref}

\usepackage[nopatch=eqnum]{microtype}
\DisableLigatures[f]{encoding = *, family = * }

\usepackage[table]{xcolor}
\usepackage{xcolor}

\usepackage{array}

\usepackage{placeins}

\newcolumntype{+}{!{\vrule width 2pt}}

\newlength\savedwidth

\raggedright
\usepackage[aboveskip=1pt,labelfont=bf,labelsep=period,justification=raggedright,singlelinecheck=off]{caption}

\makeatletter
\renewcommand{\@biblabel}[1]{\quad#1.}
\makeatother

\usepackage{lastpage,fancyhdr,graphicx}
\usepackage{epstopdf}
\fancyheadoffset[L]{2.25in}
\fancyfootoffset[L]{2.25in}
\begin{document}
\vspace*{0.2in}

\begin{flushleft}
{\Large 
\textbf{Empirical mode decomposition and interpretable machine learning for preterm birth classification from electrohysterography} % Please use "sentence case" for title and headings (capitalize only the first word in a title (or heading), the first word in a subtitle (or subheading), and any proper nouns).
}
\newline
\\
Umesha Tilakarathna\textsuperscript{1,2\Yinyang},
Senith Jayakody\textsuperscript{1,2\Yinyang,*},
Kalana Jayasooriya\textsuperscript{1},
Roshan Godaliyadda\textsuperscript{1,3},
Parakrama Ekanayake\textsuperscript{1,3},
Isuru Nawinne\textsuperscript{4},
Chathura Rathnayake\textsuperscript{5}
\\
\bigskip
\textbf{1} Multidisciplinary AI Research Centre, University of Peradeniya, Peradeniya 20400, Sri Lanka
\\
\textbf{2} Farbe Technologies (Pvt) Ltd, Peradeniya 20400, Sri Lanka
\\
\textbf{3} Department of Electrical and Electronic Engineering, University of Peradeniya, Peradeniya 20400, Sri Lanka
\\
\textbf{4} Department of Computer Engineering, University of Peradeniya, Peradeniya 20400, Sri Lanka
\\
\textbf{5} Department of Obstetrics and Gynaecology, University of Peradeniya, Peradeniya 20400, Sri Lanka
\\
\bigskip

% Insert additional author notes using the symbols described below. Insert symbol callouts after author names as necessary.
% 
% Remove or comment out the author notes below if they aren't used.
%
% Primary Equal Contribution Note
\Yinyang These authors contributed equally to this work.

% Additional Equal Contribution Note
% Also use this double-dagger symbol for special authorship notes, such as senior authorship.
% \ddag These authors also contributed equally to this work.

% \textcurrency Current Address: Dept/Program/Center, Institution Name, City, State, Country % change symbol to "\textcurrency a" if more than one current address note
% \textcurrency b Insert second current address 
% \textcurrency c Insert third current address

% \dag Deceased

% Group/Consortium Author Note
% \textpilcrow Membership list can be found in the Acknowledgments section.

% Use the asterisk to denote corresponding authorship and provide email address in note below.
* senith@eng.pdn.ac.lk

\end{flushleft}

% For PLOS Medicine research article authors, please structure your abstract
% with "Background", "Method and Findings" and "Conclusion" sections per
% journal requirements.

% For PLOS Neglected Tropical Diseases research article authors, please
% structure your abstract with "Background", "Methodology", "Findings", and
% "Conclusion" sections per journal requirements.
%
\section*{Abstract}
Preterm birth (PTB) remains a major global health problem, and reliable non-invasive risk assessment remains difficult. Electrohysterography (EHG) records uterine electrical activity from the maternal abdomen and is a candidate modality for PTB assessment, but reported performance on public EHG datasets varies widely and can be inflated when segments from the same recording are divided between training and validation folds. This study used empirical mode decomposition (EMD) to derive signal-adaptive oscillatory representations of filtered EHG signals before feature extraction, with the aim of evaluating term-versus-preterm classification. In a retrospective secondary analysis of 26 recordings, 13 preterm and 13 term, from the public 2018 Term-Preterm EHG Dataset with Tocogram, we compared dataset-provided annotated intervals with non-overlapping fixed 3-minute windows and evaluated the first four IMFs. Fourteen features spanning burst/peak morphology, temporal-energy descriptors, and entropy were extracted from each of three EHG channels, producing 42 features per segment, and evaluated using nine classifiers under repeated five-fold recording-grouped cross-validation with recording-level aggregation. IMF1 achieved the strongest mean performance among the evaluated modes. With fixed 3-minute IMF1 features, Random Forest achieved mean accuracy of 0.8308, F1 of 0.7969, balanced accuracy of 0.8308, MCC of 0.6998, ROC-AUC of 0.8157, and AP of 0.8877. Relative to the same features from the filtered time-domain signal, IMF1 yielded numerically higher values for all evaluated performance metrics, including accuracy (0.7846 to 0.8308), MCC (0.6114 to 0.6998), ROC-AUC (0.8073 to 0.8157), and AP (0.8773 to 0.8877). Preterm recordings showed smaller, more regularly spaced peak-like events, lower temporal-energy measures, and higher entropy; temporal-energy descriptors nevertheless had the largest grouped permutation importance. These findings support the methodological promise of IMF1-based EHG classification, while larger independent cohorts with explicit participant linkage are required before clinical interpretation.

% Please keep the Author Summary between 150 and 200 words. Use first person.
% PLOS ONE, PLOS Biology, PLOS Global Public Health, PLOS Mental Health, and PLOS Water authors please skip this step. Author Summary is not valid for submissions to these journals.

% For PLOS Medicine authors, please structure your author summary wghhhhhhhhhhhhhhith answers to the following questions:
% Why was this study done?
% What did the researchers do and find?
% What do these findings mean?
%
% \section*{Author summary}

% \linenumbers

% Use "Eq" instead of "Equation" for equation citations.

\section*{Introduction}
Preterm birth (PTB), defined as birth before 37 completed weeks of gestation, remains a major global health problem. An estimated 13.4 million babies (approximately one in ten live births) were born preterm in 2020 \cite{Ohuma2023,WHO_preterm_2023}. Although global under-five mortality has declined substantially over recent decades, neonatal mortality has declined more slowly than mortality among older infants and children, and neonatal deaths now account for nearly half of all under-five deaths \cite{WHO_newborn_2024,UNICEF_neonatal_2026}. Further reductions in child mortality therefore increasingly depend on improving outcomes around birth and during the neonatal period. Preterm birth is central to this burden: its complications were responsible for approximately 0.9 million deaths in 2019 and remain the leading cause of death among children younger than five years worldwide \cite{WHO_preterm_2023}. The burden is particularly high in low- and middle-income regions, including southern Asia and sub-Saharan Africa, where access to timely monitoring, referral, and neonatal care may be limited \cite{WHO_newborn_2024,WHO_preterm_2023}. Together, these estimates establish PTB as a major maternal and newborn health priority.

The impact of PTB extends beyond early mortality and includes substantial short- and long-term morbidity. Infants born preterm, particularly those born very early, are at increased risk of respiratory difficulties, feeding problems, cerebral palsy, developmental delay, vision impairment, hearing impairment, and other long-term complications \cite{CDC_preterm_2024}. These outcomes can impose lifelong consequences for the child, emotional and financial strain on families, and substantial demands on healthcare systems. Therefore, reducing the burden of PTB is not only a neonatal survival priority, but also a major concern in terms of long-term morbidity, public health, socioeconomic burden, and family welfare.

Timely identification of women at increased risk of preterm delivery may enable closer surveillance and referral. When preterm labour or delivery is clinically suspected within an actionable gestational window, guideline-based pathways may include administration of antenatal corticosteroids for fetal lung maturation, tocolysis when appropriate, and neonatal transfer or care planning \cite{who2022tocolytic,ACOG2017_corticosteroids,NICE_NG25}. Improved risk stratification should therefore be viewed as a potential adjunct to established clinical assessment, not as a stand-alone diagnostic or treatment trigger.

\subsection*{Challenges in practical PTB risk assessment}

Several approaches have been investigated or used for assessing PTB risk or suspected preterm labour, including clinical history, obstetric risk factors, cervical length measurement, fetal fibronectin testing, biomarker tests, uterine activity monitoring, and imaging-based assessment. These methods provide useful clinical information, but accurate, generalizable, and practically deployable PTB risk assessment remains challenging. Clinical risk factors alone are often insufficient because PTB is multifactorial and can occur in women without obvious prior risk. Cervical length measurement and fetal fibronectin testing are clinically relevant, particularly in symptomatic or high-risk settings, but their use depends on gestational age, clinical presentation, measurement setting, operator expertise, and intended prediction window \cite{NICE_NG25,SonMiller2017}. 
Current guidelines therefore use these tests within well defined clinical pathways rather than as universal screening tools \cite{NICE_NG25}.

Electrohysterography (EHG) records myometrial electrical activity non-invasively using electrodes on the maternal abdomen and is closely related to uterine contractile activity \cite{GarfieldManer2007,Lucovnik2011,Buhimschi1997}. External tocodynamometry, by comparison, measures mechanical deformation at the abdominal surface. EHG therefore provides a more direct view of uterine electrophysiology and may support repeated, low-burden measurement. However, EHG is not established as a routine tool for PTB risk stratification. Variability in acquisition and preprocessing, limited external validation, and uncertainty about the clinical interpretation and incremental value of EHG-derived features remain barriers to translation \cite{GarciaCasado2018_EHGReview}. These gaps motivate reproducible and physiologically interpretable EHG analysis.

\subsection*{Physiological basis of uterine electrical activity}
The myometrium is the smooth-muscle layer of the uterus. Myometrial electrical activity and its propagation depend on cellular excitability and intercellular coupling, including coupling through gap junctions \cite{GarfieldSims1977,Aguilar2010}. For much of pregnancy, the uterus is relatively quiescent and myometrial electrical activity is weakly coordinated. As labour approaches, changes in contraction-associated proteins, ion channels, receptors, and connexin-43 gap junctions increase excitability, electrical propagation, and coordination of contractions \cite{Aguilar2010, garfield1998control}. This transition is regulated through interacting endocrine, paracrine, inflammatory, and mechanical pathways \cite{garfield1998control, garfield1993control}.

These preparatory changes are reflected in uterine electrical activity before and during labour. Surface uterine electromyography, recorded from the maternal abdomen as EHG, has therefore been investigated for characterizing the transition from relative quiescence to the electrically active labour state \cite{garfield1998control}. Increased uterine electrical activity has been reported during both term and preterm labour \cite{wolfs1979electromyographic, csapo1981force}, consistent with increased contractile activity.

These observations support the hypothesis that myometrial activation associated with labour may occur prematurely in pregnancies ending preterm and may produce detectable changes in surface EHG. They therefore provide a physiological rationale for evaluating EHG features related to burst morphology, temporal energy, and signal complexity as candidate markers for retrospective term-versus-preterm classification. Prospective studies with defined prediction horizons are required to determine whether such features can support clinical risk stratification.

\subsection*{EHG signal characteristics and bursts}

Surface EHG records the abdominal manifestation of myometrial electrical activity. Although maternal tissues attenuate the signal, burst-like activity associated with contractions remains detectable from abdominal recordings \cite{Buhimschi1997,GarfieldManer2007}. Much of the spectral power associated with uterine contractions is reported below approximately 1 Hz \cite{Devedeux1993}; however, this is not a physiological or analytical cutoff, because faster uterine components may extend above 1 Hz and non-uterine sources can overlap the recorded band. The present analysis therefore used the dataset-provided signals filtered from 0.08 to 5.0 Hz and did not impose an additional low-pass cutoff at 1 Hz or assign a predefined frequency band to any IMF. Because EMD is signal-adaptive, the spectral content of a given IMF may vary among segments and recordings.

Uterine EMG/EHG bursts are time-localized episodes of electrical activity rather than a single frequency band. Their occurrence, morphology, internal oscillatory content, and energy have been investigated as indicators of the transition from relative uterine quiescence toward labour. Li et al. reported that uterine EMG burst power increased with cervical dilation and reached higher values as labour progressed; frequency-related parameters increased earlier, whereas power-related parameters became stronger later in labour \cite{li2022uterine}. Buhimschi et al. similarly reported that uterine electrical activity was minimal during much of pregnancy, but appeared as bursts that became more frequent and larger in amplitude during term and preterm labour \cite{Buhimschi1997}. These labour-associated bursts correlated with contractions, patient-reported pain or pressure, and changes in intrauterine pressure \cite{Buhimschi1997}.

Evidence from threatened preterm labour further supports evaluating burst morphology. Mas-Cabo et al. reported an increasing trend in EHG burst amplitude as labour approached, with representative recordings showing larger bursts among women delivering within seven days than among those delivering later \cite{mascabo2019uterine}. However, amplitude alone was not a robust discriminator. Burst amplitude should therefore be evaluated alongside event timing, organization, energy, and complexity rather than assumed to increase uniformly in every term–preterm comparison.

Taken together, these findings support analysing EHG bursts as multicomponent events rather than relying on a single frequency interval or amplitude measure. The 14 feature types evaluated per channel therefore describe complementary properties of the signal: peak and burst descriptors capture event rate, spacing, amplitude, width, and grouping; temporal-energy descriptors capture waveform variation and energy; and entropy descriptors capture regularity and complexity. This framework allows term–preterm separation to arise from a joint feature pattern in which different descriptors may vary in different directions, rather than assuming that all features increase as labour approaches.

\subsection*{Computational approaches and methodological pitfalls}
Recent studies have applied diverse computational approaches to EHG-based term–preterm classification, including handcrafted temporal and spectral descriptors, selection from larger linear and nonlinear feature pools, time–frequency and entropy representations, multichannel propagation and synchronization measures, and deep-learning models \cite{FeleZorz2008, Jager2018_TPEHGT, Fergus2013_EHGML, Alamedine2013_EHGFeatureSelection, RomeroMorales2023, Goldsztejn2023_EHGDeepLearning, Fischer2023_EHGDeepLearning}. These studies demonstrate the potential of EHG, but also show that performance depends strongly on signal representation, feature construction, classifier choice, and validation design. In many pipelines, the primary emphasis is on selecting a high-performing feature subset or learning a predictive representation, while the physiological meaning of the retained signal properties and their contribution to the final model decision receive less systematic attention \cite{Pirnar2024}. This limits assessment of whether strong performance reflects plausible uterine signal characteristics or dataset-specific structure. There is therefore value in frameworks that retain traceable signal descriptors and examine both class-dependent feature behavior and model-level feature-family importance.

Reported performance is also highly sensitive to data partitioning. Applying oversampling before dividing data into training and validation sets can introduce information leakage and produce optimistic performance estimates \cite{Vandewiele2021}. For segmented biomedical signals, a separate preventable error occurs when windows or contraction intervals from the same recording appear in both training and validation data, allowing recording-specific structure to be shared across the split. All segments derived from one recording should therefore be assigned exclusively to either training or validation within each fold, and final predictions and metrics should be computed at the recording level.

Within this setting, empirical mode decomposition (EMD) is well suited to intermittent, nonstationary EHG because it derives intrinsic mode functions (IMFs) from the local extrema and characteristic time scales of each input signal rather than imposing a fixed global basis or predefined frequency boundaries \cite{Huang1998_EMD}. Fourier- and wavelet-based analyses use predefined analysis functions and selected frequency bands or scales, within which physiological activity and artifacts occupying the same range may remain combined. EMD instead organizes local oscillatory content into data-dependent modes, providing a plausible representation for extracting temporally localized morphology, energy, and complexity descriptors. This adaptivity does not guarantee denoising or assign a fixed physiological meaning to an IMF; the relative usefulness of the modes must therefore be established empirically.

To retain interpretability, the extracted features were chosen to represent traceable signal properties. Peak and burst features describe event timing, morphology, and grouping; temporal-energy features describe waveform variation and energy; and entropy features describe signal regularity and complexity. These descriptors may not change in the same direction, and their relevance can be assessed at both the univariate descriptive and multivariate model levels. Accordingly, class-dependent differences in a feature should be interpreted alongside its contribution to the multivariate model, rather than as evidence that the feature is independently predictive or represents a standalone physiological biomarker.

\subsection*{Study contributions}

Against this background, the study makes four bounded methodological contributions:

\begin{enumerate}
    \item \textbf{Signal-adaptive representation with a matched comparator.}  
    The first four EMD-derived IMFs are evaluated under the same feature and recording-level classification framework, after which the selected IMF is compared with identical features extracted from the filtered time-domain signal. This isolates the effect of signal representation.

    \item \textbf{Prevention of recording-level segment leakage.}
    All segments from a recording remain within one cross-validation fold, no oversampling is used, and segment-level classification scores are aggregated into one recording-level score. This addresses the direct, preventable leakage pathway created when segments from one recording cross the split.

    \item \textbf{Direct comparison of segmentation strategies.}
    Dataset-provided annotated intervals, comprising contraction and dummy (non-contraction) intervals, are compared with non-overlapping fixed three-minute windows under the same framework. The fixed-window strategy yields the highest observed mean recording-level performance among the evaluated segmentation settings, indicating that useful classification information is not confined to annotated contractions while also avoiding dependence on manual interval selection.

    \item \textbf{Interpretable feature-family evidence.}
    Fourteen feature types per channel are examined through recording-level effect directions and grouped permutation importance. The effect directions identify a coherent preterm-associated pattern of more frequent, smaller, and more regularly spaced peak-like events, lower waveform magnitude/energy, and higher entropy. Temporal-energy descriptors contribute most strongly to the selected Random Forest models, burst and peak descriptors provide complementary information, and entropy provides little consistent unique importance. This distinguishes descriptive feature patterns from their contribution to model predictions, without implying that any individual feature or the proposed multi-scale organization is a causal biomarker.
    
\end{enumerate}

Among the evaluated configurations, fixed three-minute IMF1 features with Random Forest produced the highest observed mean performance. Relative to the matched filtered time-domain representation, IMF1 produced numerically higher mean values across all six reported metrics, although the differences in ROC-AUC and AP were modest.

\section*{Materials and methods}

\begin{figure}[!h]
\centering
\includegraphics[width=\textwidth]{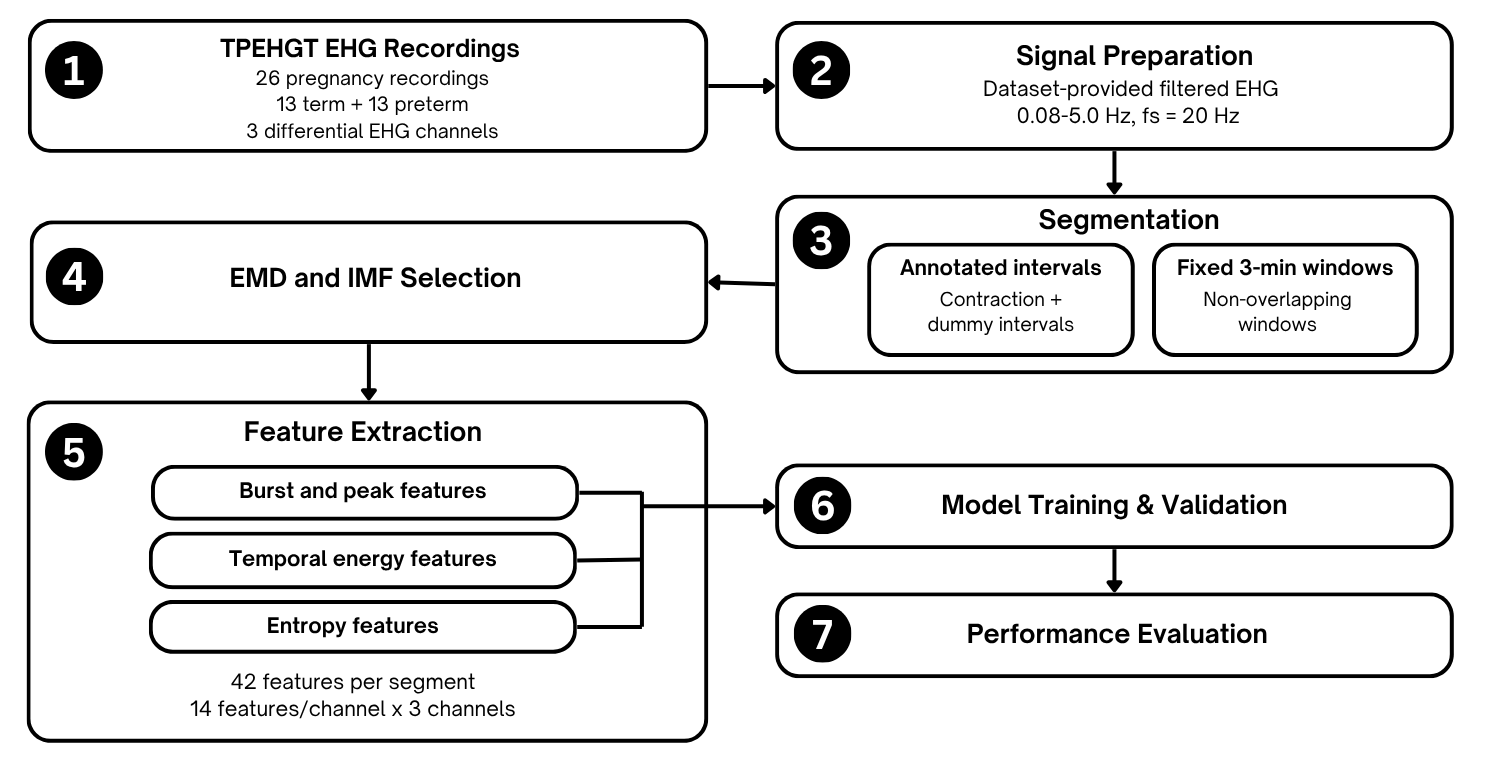}
\caption{\textbf{Overview of the proposed EHG-based recording-level term-versus-preterm classification framework.} Dataset-provided filtered EHG recordings were segmented using annotated-interval and non-overlapping fixed 3-minute strategies. Each segment was decomposed using empirical mode decomposition (EMD), and the first four intrinsic mode functions (IMF1-IMF4) were evaluated under the same feature extraction and classification framework. IMF1 was selected based on the resulting recording-level performance and used for the subsequent analyses. Burst and peak, temporal-energy, and entropy features were extracted from the selected representation and used to train machine-learning classifiers. Segment-level classification scores were aggregated using maximum-score aggregation to obtain one recording-level score for classification.}
\label{fig:pipeline}
\end{figure}

The proposed framework is illustrated in Fig.~\ref{fig:pipeline}. The analysis consisted of signal segmentation, EMD-based decomposition, IMF selection, feature extraction, classification, and recording-level aggregation. The following subsections describe the dataset, preprocessing, feature extraction, model training, and evaluation protocol.

\subsection*{Dataset}

The publicly available Term-Preterm EHG Dataset with Tocogram (TPEHGT) \cite{Jager2018_TPEHGT,Pollard2026PhysioNet} was used in this study. The dataset was developed at the Faculty of Computer and Information Science, University of Ljubljana, Slovenia, with recordings collected at the Department of Obstetrics and Gynecology, University Medical Center Ljubljana.

The dataset comprises 31 approximately 30-minute uterine recordings. Of these, 13 recordings were obtained from eight pregnancies that ended in spontaneous preterm delivery and 13 recordings from ten pregnancies that ended in spontaneous term delivery. The mean gestational ages at delivery were $33.7 \pm 1.97$ weeks for the preterm group and $38.1 \pm 1.04$ weeks for the term group. The pregnancy recordings were acquired during routine antenatal visits at a pooled mean gestational age of $30.2 \pm 2.76$ weeks. The remaining five recordings were obtained from non-pregnant women. The preterm recordings contain 47 annotated contraction intervals and 47 annotated dummy (non-contraction) intervals, whereas the term recordings contain 53 annotated contraction intervals and 53 annotated dummy intervals.

Recordings were acquired using four abdominal surface electrodes arranged in two horizontal rows symmetrically above and below the navel, with 7~cm spacing. From these electrodes, the dataset provides three differential EHG signals, $S_1 = E_2-E_1$, $S_2 = E_2-E_3$, and $S_3 = E_4-E_3$, together with a simultaneously recorded tocogram (TOCO) signal obtained using an external tocodynamometer.

The present analysis included the 26 pregnancy recordings corresponding to term and preterm deliveries and excluded the five recordings from non-pregnant women. These 26 recordings originated from 18 pregnancies. Because the public release does not provide an explicit recording-to-pregnancy mapping suitable for pregnancy-level grouping, cross-validation was performed using the recording as the grouping unit. Consequently, possible dependence between multiple recordings originating from the same pregnancy could not be controlled.

Only the three differential EHG signals were used as classifier inputs. The TOCO signal itself was not used as a predictor because the proposed features were intended to characterize uterine electrical rather than mechanical activity. However, TOCO contributed indirectly to the annotation-guided analysis because the dataset-provided contraction annotations were defined by the dataset creators with reference to TOCO deflections together with accompanying EHG activity.

The original data collection was approved by the National Medical Ethics Committee of the Republic of Slovenia (approvals No.~32/01/97 and No.~108/09/09), and written informed consent was obtained from participants as reported by the dataset creators. The present study involved secondary analysis of publicly available de-identified data and no new participant recruitment or data collection.

\subsection*{Preprocessing and segmentation}

The three differential EHG signals were sampled at 20~Hz and were segmented using two temporal strategies before EMD and feature extraction.

\subsubsection*{Signal preparation}

The dataset-provided filtered EHG signals were used as inputs to the proposed framework. In the TPEHGT release, each differential EHG signal is provided in both unfiltered and filtered form. The filtered signals were obtained by the dataset creators using a four-pole bidirectional Butterworth band-pass filter with cutoff frequencies of 0.08 and 5.0~Hz. Accordingly, no additional band-pass filtering was applied in the present study before segmentation or EMD.

EMD was applied directly to the dataset-provided filtered EHG segments, for which very-low-frequency components below 0.08~Hz and components above 5.0~Hz had already been attenuated by the dataset preprocessing. No fixed-window segment or dataset-provided annotated interval was removed by an additional manual or automated artifact-rejection rule in the present analysis.

\subsubsection*{Segmentation strategies}

Two segmentation strategies were used to examine how temporal segmentation influenced recording-level term-versus-preterm classification.

\paragraph{Annotated-interval segmentation}

The annotation-guided strategy used both types of manually annotated intervals provided with the TPEHGT dataset: contraction intervals and dummy (non-contraction) intervals. The preterm recordings contain 47 contraction and 47 dummy intervals, whereas the term recordings contain 53 contraction and 53 dummy intervals. Thus, 200 annotated intervals were included across the 26 pregnancy recordings. Each annotated interval was treated as an individual segment for subsequent EMD and feature extraction. The number of annotated intervals varied among recordings according to the dataset-provided annotations.

\paragraph{Fixed 3-minute segmentation}

Each recording was divided into consecutive, non-overlapping complete 3-minute windows (180~s). At the sampling rate of 20~Hz, each complete window contained 3600 samples. Any trailing portion shorter than 180~s was excluded rather than padded or treated as a shorter segment. This procedure yielded 249 fixed-length segments across the 26 pregnancy recordings: 15 recordings contributed ten complete windows and 11 recordings contributed nine. Unlike the annotation-guided strategy, fixed-window segmentation did not depend on manually defined interval annotations and sampled the recordings using a uniform temporal representation.

\begin{figure}[!htbp]
\centering
\includegraphics[width=\textwidth]{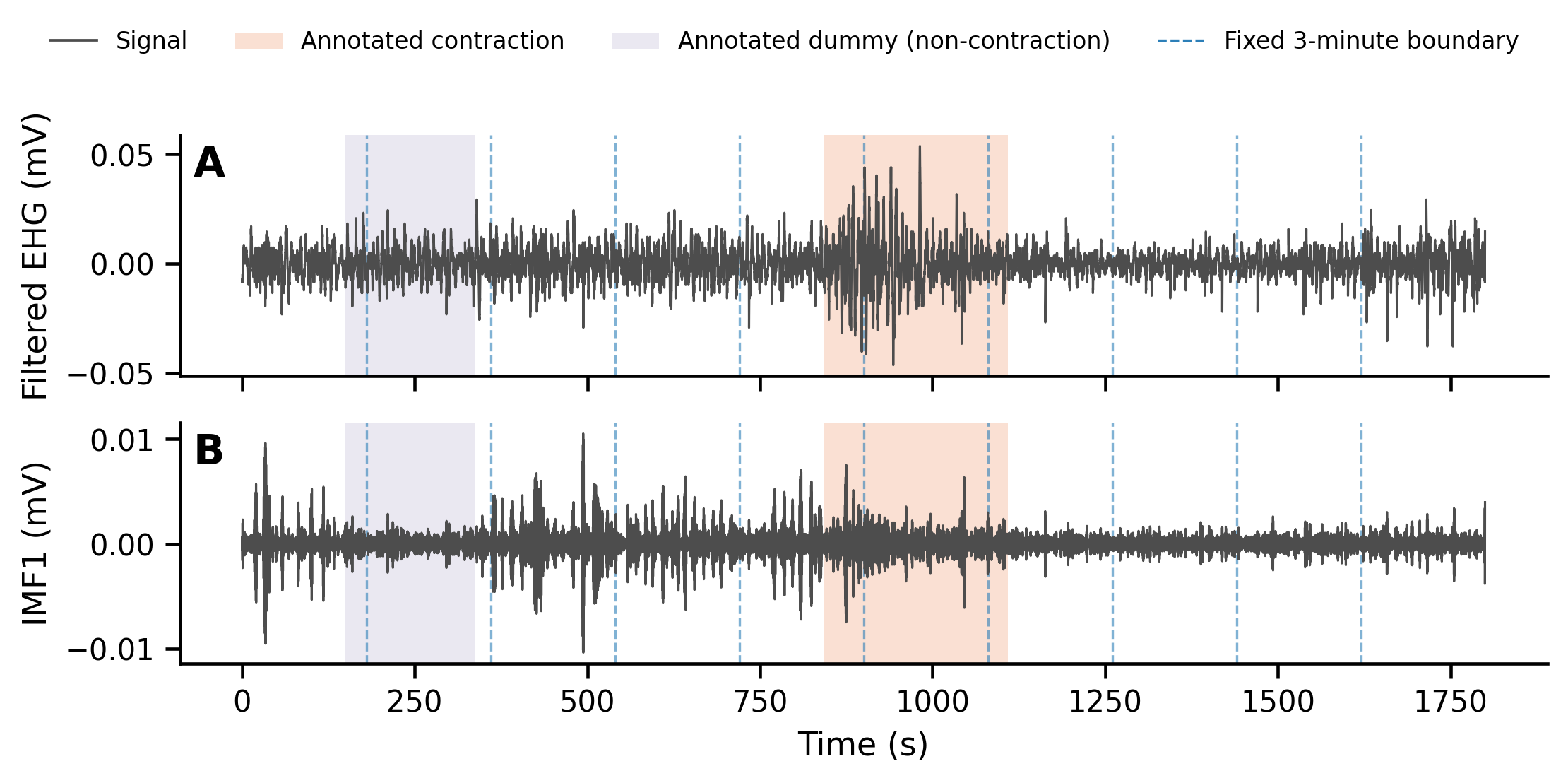}
\caption{\textbf{Comparison of annotated-interval and fixed 3-minute segmentation on a representative recording.} (A)~Filtered EHG signal from one differential channel, with dataset-provided contraction intervals and dummy (non-contraction) intervals shown as separately shaded regions and fixed 3-minute window boundaries shown as dashed vertical lines. (B)~An IMF1 representation of the same recording shown only to illustrate how the segmentation schemes relate to oscillatory EHG activity. The annotated-interval strategy uses both contraction and dummy intervals, whereas the fixed-window strategy uses consecutive non-overlapping complete 3-minute windows irrespective of the annotations. For the analytical pipeline, segmentation was performed before EMD and each resulting segment was decomposed independently.}
\label{fig:segmentation}
\end{figure}

The qualitative difference between the two strategies is illustrated in Fig.~\ref{fig:segmentation}. The annotated-interval strategy retains the dataset-provided contraction and dummy intervals, whereas fixed-window segmentation divides the available recording duration into consecutive complete 3-minute windows irrespective of the interval annotations. In the analytical pipeline, segmentation was performed before EMD, and each resulting segment was decomposed independently.

\subsubsection*{Recording-level data separation}

To prevent direct leakage between segments derived from the same recording, cross-validation partitions were constructed at the recording level. For each split, one class label was associated with each of the 26 recordings, and stratification was performed on these recording-level entries. The resulting training and validation recording identifiers were then mapped back to their constituent segments, ensuring that all segments from a given recording remained entirely within either the training or validation set in that fold. No oversampling was applied.

Grouping could not be performed at the pregnancy level because the public TPEHGT release does not provide an explicit recording-to-pregnancy mapping suitable for constructing pregnancy-grouped folds. Consequently, possible dependence between multiple recordings originating from the same pregnancy cannot be excluded. The complete repeated cross-validation and recording-level evaluation procedure is described in the Evaluation metrics subsection.

\subsection*{Empirical mode decomposition}

Empirical mode decomposition (EMD) was used to obtain data-adaptive oscillatory components from the filtered EHG signals~\cite{Huang1998_EMD}. Unlike decomposition methods based on predefined basis functions, EMD represents a signal according to its local characteristic time scales and is therefore well suited to nonlinear and nonstationary signals. This property is particularly relevant to EHG, in which uterine electrical activity is intermittent and its oscillatory structure can vary over time.

For a filtered EHG segment $s[n]$, EMD iteratively identifies the local maxima and minima and interpolates them to form upper and lower envelopes. The mean of these envelopes is removed from the signal through the sifting process until the resulting component satisfies the conditions of an intrinsic mode function (IMF). After extraction of an IMF, the same procedure is applied to the remaining signal to obtain progressively slower oscillatory components. The resulting decomposition can be expressed as

\begin{equation}
s[n] = \sum_{k=1}^{K} c_k[n] + r[n],
\label{eq:emd}
\end{equation}

where $c_k[n]$ denotes the $k$th IMF and $r[n]$ is the residual component. Lower-order IMFs generally represent faster local oscillations, whereas higher-order IMFs describe progressively slower variations in the signal.

EMD was applied independently to each EHG channel within each segmented interval using the PyEMD implementation (EMD-signal v1.9.0) with its standard sifting and stopping criteria. Decomposition was limited to the first four IMFs required for the present analysis. The residual returned by EMD was kept separate from the IMFs and was not used as a feature-source representation for classification.

\begin{figure}[!htbp]
\centering
\includegraphics[width=\textwidth]{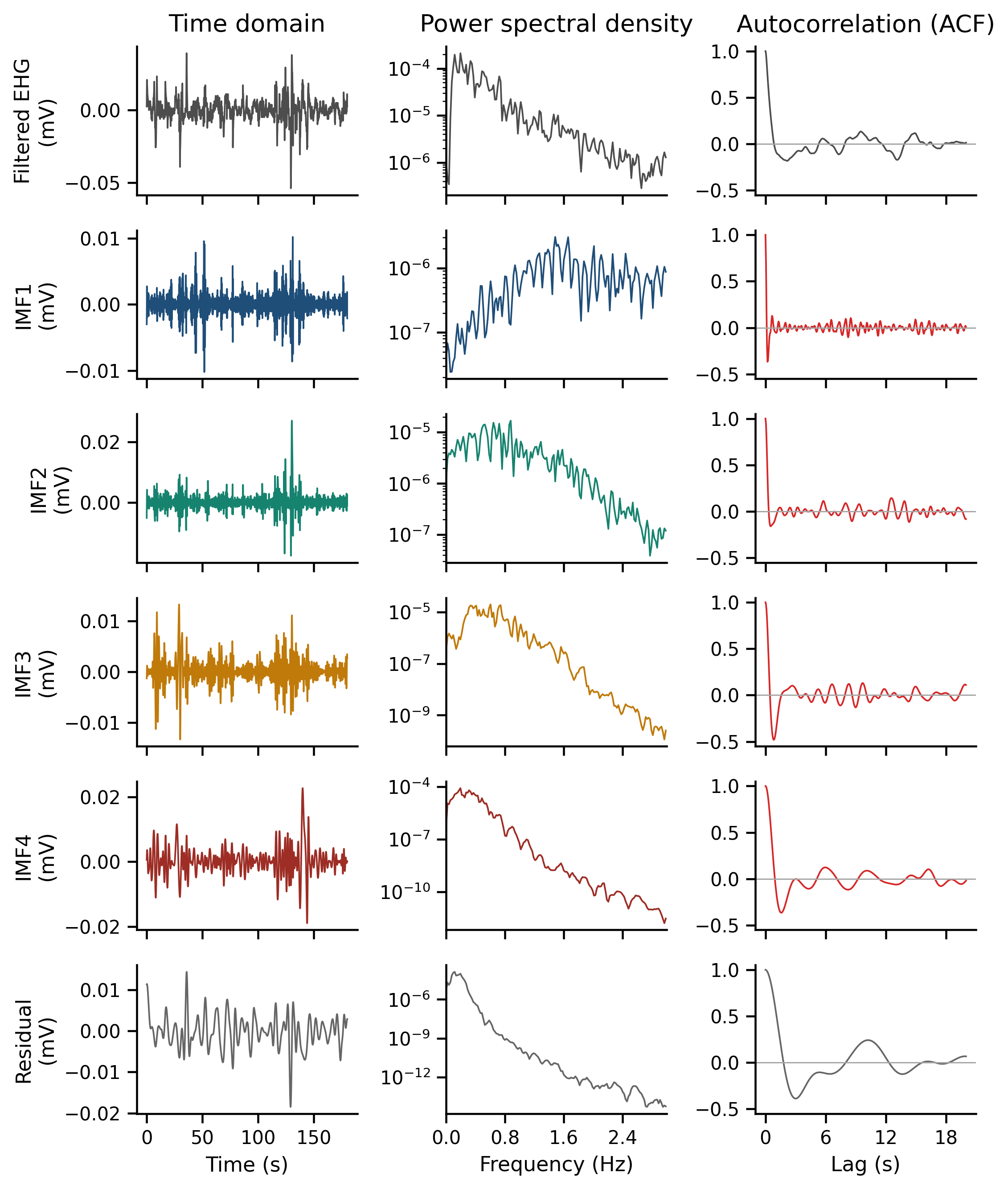}
\caption{\textbf{Representative empirical mode decomposition of a filtered EHG segment.} The dataset-provided filtered EHG segment was decomposed into IMF1-IMF4 and a residual component. For each component, the time-domain waveform, PSD, and autocorrelation function are shown. The example illustrates the progression from faster local oscillations in the lower-order IMFs to slower variation in the higher-order IMFs and residual.}
\label{fig:imf_acf_psd}
\end{figure}

Fig.~\ref{fig:imf_acf_psd} illustrates the first four IMFs and residual obtained from a representative filtered EHG segment, together with their power spectral densities (PSDs) and autocorrelation functions.

No fixed physiological frequency band was assigned to an individual IMF. Because EMD is data-adaptive, the spectral content of a given IMF can vary across segments and recordings. The first four IMFs were therefore compared empirically under the same downstream feature-extraction and classification framework, and their spectral characteristics were examined after decomposition.

\subsection*{Selection of the intrinsic mode function}

The first four IMFs were evaluated separately to examine which mode provided the strongest representation for the selected feature framework. For each IMF, the same 42-dimensional channel-wise feature set, classifier set, recording-level aggregation, and cross-validation procedure were used. No IMF was selected a priori on the basis of its nominal order or an assumed physiological frequency interval.

The comparison was performed within the fixed 3-minute segmentation setting. For each IMF, the classifier with the highest mean recording-level balanced accuracy was identified, and the resulting IMF-level performances were compared using the same criterion. IMF1 yielded the highest mean balanced accuracy among IMF1--IMF4 and was therefore carried forward for the subsequent feature, segmentation, and model analyses.

Because IMF and classifier selection were performed using the same dataset on which their cross-validated performance was summarized, this comparison is considered an exploratory within-dataset selection rather than an independent validation of IMF1. The resulting performance estimates may therefore contain some selection optimism and require confirmation in an independent dataset.

\begin{figure}[!htbp]
\centering
\includegraphics[width=0.85\textwidth]{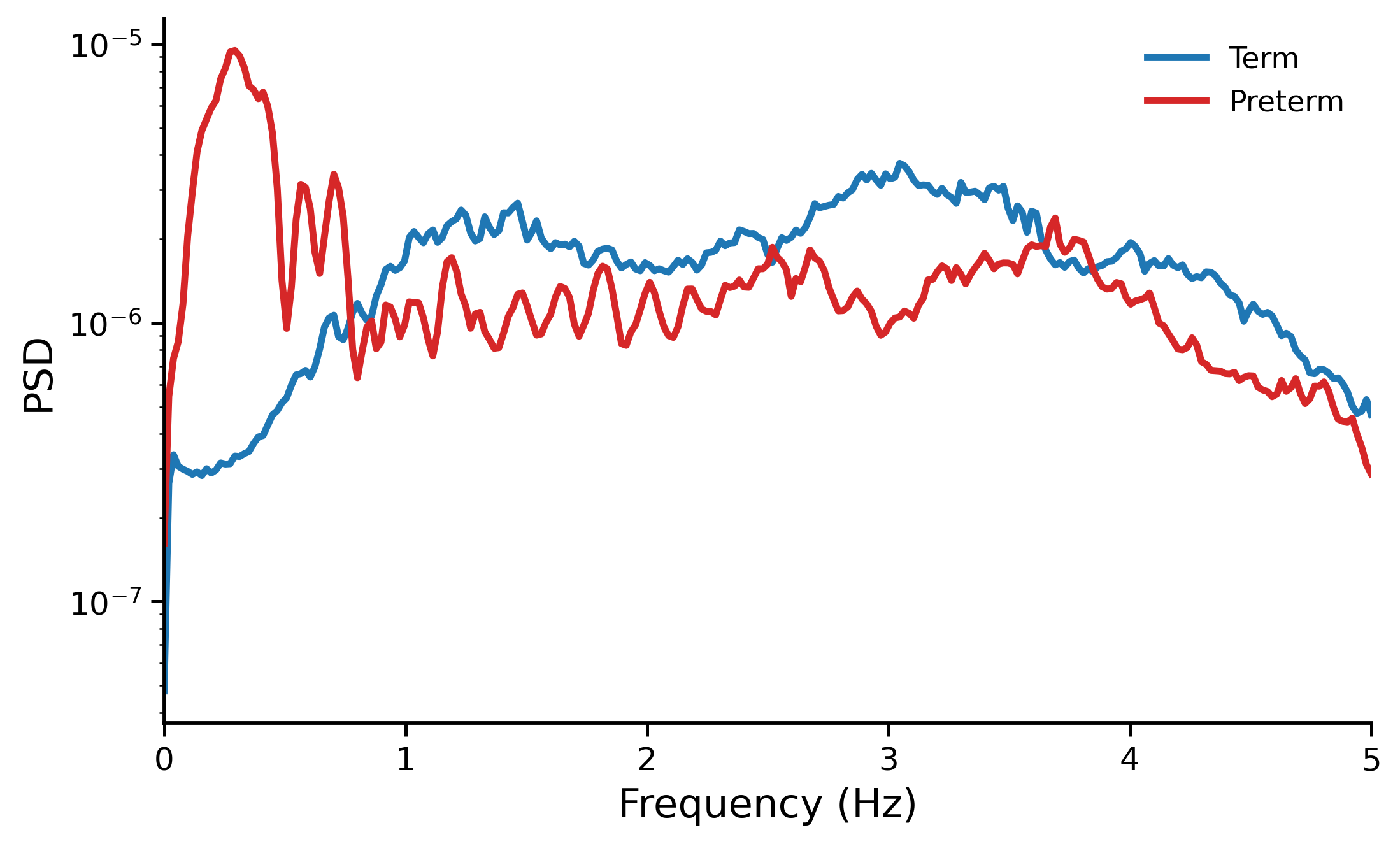}
\caption{\textbf{Average IMF1 PSD by class.} Welch PSDs were estimated from IMF1 components extracted from fixed 3-minute filtered EHG segments using a Hann window of length 1024 samples with 50\% overlap. For each recording, PSDs were averaged across segments and across the three EHG channels, and the resulting recording-level PSDs were averaged within each class. Term recordings are shown in blue and preterm recordings in red. The y-axis is log scaled and the frequency range is shown from 0 to 5~Hz. The curves show class-dependent spectral differences, with preterm recordings showing higher IMF1 power mainly in the lower frequency portion of the displayed range.}
\label{fig:psd}
\end{figure}

After this empirical selection, the spectral distribution of IMF1 was examined descriptively rather than used as an additional selection criterion. Welch PSDs~\cite{Welch1967} were averaged within each recording and subsequently within each class, as shown in Fig.~\ref{fig:psd}. For notation, the selected IMF1 sequence is denoted by $x[n]=c_1[n]$ in the remainder of the Methods section.

\subsection*{Feature extraction}

Fourteen feature types were extracted independently from each of the three EHG channels, and the channel-wise values were concatenated to form a 42-dimensional feature vector for each segment. The features were organized into three families: eight peak and burst descriptors, three temporal-energy descriptors, and three entropy descriptors. These families were selected to characterize complementary aspects of event timing and morphology, waveform variation, typical magnitude, and local energy, and nonlinear signal complexity.

For the EMD-based analysis, the features were calculated from the empirically selected IMF1 representation, denoted by $x[n]=c_1[n]$. For the matched filtered time-domain comparator, the same feature definitions, channel concatenation, and segmentation-specific feature parameters were applied directly to the corresponding dataset-filtered EHG segments; only the input signal representation differed. This matched design enabled the contribution of the EMD-derived representation to be examined without changing the feature set or downstream classification framework.

The selected feature families were motivated by the burst-like and nonstationary characteristics of uterine electrical activity described in the Introduction. Peak and burst descriptors characterize the timing, density, morphology, and grouping of algorithmically detected peak-like events; temporal-energy descriptors summarize waveform variation and energy within the analyzed representation; and entropy descriptors quantify regularity and complexity~\cite{GarfieldManer2007, garfield1998control,Lammers2013}.

\subsubsection*{Burst and peak features}

Peak and burst-related features were calculated from the magnitude of the analyzed signal representation,

\begin{equation}
z[n] = |x[n]|.
\end{equation}

\begin{figure}[!htbp]
\centering
\includegraphics[width=\textwidth]{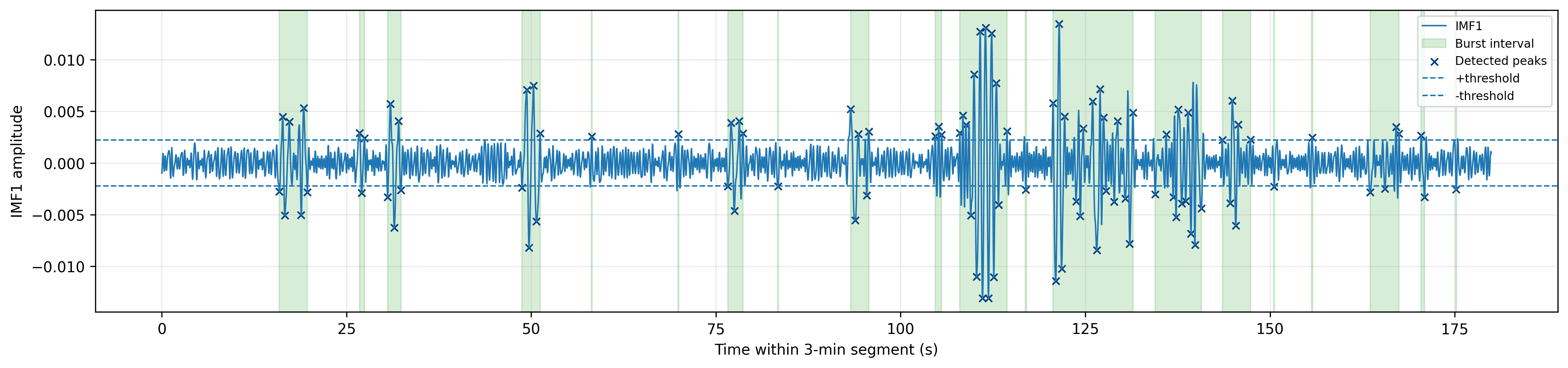}
\caption{\textbf{Representative peak and burst-like event detection on the selected IMF1 component.} Detected peak-like events were identified from the IMF1 magnitude $|x[n]|$ and are marked at their corresponding locations on the IMF1 waveform. Dashed horizontal lines indicate the magnitude-based detection threshold, shown at $\pm T_{\mathrm{peak}}$ for visualization, and shaded regions indicate burst-like groups formed by temporally adjacent detected peaks.}
\label{fig:peak_burst}
\end{figure}

A local maximum was retained as a peak when its magnitude exceeded the robust adaptive threshold

\begin{equation}
T_{\mathrm{peak}}
=
\operatorname{median}(z)
+
\mathrm{THK}\,\operatorname{MAD}(z),
\label{eq:peak_threshold}
\end{equation}

where THK is the threshold multiplier and MAD denotes the median absolute deviation. Detected peaks were required to be separated by at least MD seconds. Consecutive peaks were assigned to the same burst-like group when their inter-peak interval was less than or equal to BT seconds; otherwise, a new group was initiated. Fig.~\ref{fig:peak_burst} illustrates this procedure.

Eight descriptors were derived from the detected peak and burst-like structure. For a segment of duration $D$ seconds containing $P$ detected peaks, peak rate was defined as

\begin{equation}
\mathrm{Peak\ Rate} = \frac{P}{D}.
\end{equation}

Peak rate therefore describes the temporal density of detected peak-like events.

For detected peak magnitudes $\{a_i\}_{i=1}^{P}$, mean peak amplitude was computed as

\begin{equation}
\mathrm{Peak\ Amp\ Mean}
=
\frac{1}{P}\sum_{i=1}^{P} a_i .
\end{equation}

The coefficient of variation of peak amplitude was calculated as

\begin{equation}
\mathrm{Peak\ Amp\ CV}
=
\frac{\sigma_a}{\mu_a+\epsilon},
\end{equation}

where $\mu_a$ and $\sigma_a$ are the mean and population standard deviation of the detected peak magnitudes, respectively.

For peak locations $\{t_i\}_{i=1}^{P}$ expressed in seconds, consecutive inter-peak intervals were defined as

\begin{equation}
\Delta t_i = t_{i+1}-t_i .
\end{equation}

The mean inter-peak interval was

\begin{equation}
\mathrm{IPI\ Mean}
=
\frac{1}{P-1}
\sum_{i=1}^{P-1}\Delta t_i ,
\end{equation}

and its coefficient of variation was

\begin{equation}
\mathrm{IPI\ CV}
=
\frac{\sigma_{\Delta t}}
{\mu_{\Delta t}+\epsilon},
\end{equation}

where $\mu_{\Delta t}$ and $\sigma_{\Delta t}$ denote the mean and population standard deviation of the inter-peak intervals.

Peak width was estimated from the magnitude signal using the width at half prominence of each detected peak. For detected widths $\{w_i\}_{i=1}^{P}$ expressed in seconds, mean peak width was defined as

\begin{equation}
\mathrm{Peak\ Width\ Mean}
=
\frac{1}{P}\sum_{i=1}^{P}w_i .
\end{equation}

If $B$ burst-like groups were identified within a segment, burst rate was defined as

\begin{equation}
\mathrm{Burst\ Rate}
=
\frac{B}{D}.
\end{equation}

Normalization by segment duration was used because the annotated intervals have variable durations, whereas the fixed-window segments have a constant duration.

Finally, if $q_b$ denotes the number of detected peaks in burst-like group $b$, the mean number of peaks per burst was computed as

\begin{equation}
\mathrm{Mean\ Peaks\ per\ Burst}
=
\frac{1}{B}\sum_{b=1}^{B}q_b .
\end{equation}

These descriptors characterize algorithmically detected peak timing, magnitude, width, spacing, and grouping within the analyzed signal representation. They were not interpreted as event-by-event physiological contraction labels, and no uniform direction of change between term and preterm recordings was assumed.

A small numerical constant of $\epsilon=10^{-12}$ was used where required to avoid numerical singularities. When no peak or burst-like event was detected, peak rate and burst rate were assigned their meaningful value of zero. Quantities that could not be estimated because an insufficient number of peaks or peak pairs was available were treated as missing values rather than as numerical zeros. Missing feature values were subsequently imputed using means estimated exclusively from the corresponding cross-validation training partition.

\subsubsection*{Temporal energy features}

Three temporal-energy descriptors were calculated from the analyzed signal representation to characterize complementary aspects of waveform variation, typical magnitude, and local energy. These descriptors were Difference Absolute Standard Deviation Value (DASDV), log detector, and mean Teager-Kaiser energy (MTKE), which have previously been used in EMD-based EHG analysis for term-preterm classification~\cite{janjarasjitt2021preterm,janjarasjitt2022comparison,mohammadifar2022prediction,mohammadifar2023prediction}. Let $x[n]$ denote the analyzed signal sequence of length $N$.

DASDV quantifies short-lag waveform variation through the root mean square of consecutive sample differences:

\begin{equation}
\mathrm{DASDV}
=
\sqrt{
\frac{1}{N-1}
\sum_{n=1}^{N-1}
\left(x[n+1]-x[n]\right)^2
}.
\end{equation}

The log detector provides a compressed measure of the typical signal magnitude and was calculated as

\begin{equation}
\mathrm{LOG}
=
\exp\left(
\frac{1}{N}
\sum_{n=1}^{N}
\ln\left(|x[n]|+\epsilon\right)
\right),
\end{equation}

where $\epsilon=10^{-12}$ was included to avoid evaluating the logarithm at zero.

The Teager-Kaiser energy operator~\cite{Kaiser1990} was defined for the interior samples as

\begin{equation}
\Psi[x[n]]
=
x[n]^2-x[n-1]x[n+1],
\end{equation}

and MTKE was computed as the arithmetic mean of the unrectified operator:

\begin{equation}
\mathrm{MTKE}
=
\frac{1}{N-2}
\sum_{n=2}^{N-1}
\Psi[x[n]].
\end{equation}

DASDV therefore characterizes short-lag waveform change, the log detector summarizes typical signal magnitude, and MTKE characterizes mean local Teager-Kaiser energy. These quantities were treated as complementary signal descriptors rather than collectively as measures of variability.

\subsubsection*{Entropy features}

Permutation entropy, sample entropy, and Shannon entropy were used to characterize complementary aspects of signal complexity and organization. Permutation entropy describes the diversity of local ordinal patterns, sample entropy characterizes the regularity or predictability of waveform patterns, and Shannon entropy summarizes the dispersion of the signal amplitude distribution. These measures were treated as signal-level descriptors rather than direct measures of global uterine synchronization.

Permutation entropy~\cite{BandtPompe2002} was calculated from ordinal patterns formed from embedding vectors

\begin{equation}
\mathbf{v}_i =
[x[i],x[i+\tau],\ldots,x[i+(m-1)\tau]].
\end{equation}

If $p_j$ denotes the probability of the $j$th ordinal pattern, permutation entropy was defined as

\begin{equation}
H_{\mathrm{PE}}
=
-\sum_{j=1}^{m!} p_j \ln(p_j),
\end{equation}

and normalized as

\begin{equation}
H_{\mathrm{PE}}^{\mathrm{norm}}
=
\frac{H_{\mathrm{PE}}}{\ln(m!)}.
\end{equation}

In this study, an embedding dimension of $m=3$ and delay $\tau=1$ were used, giving normalization by $\ln(3!)$.

Sample entropy~\cite{RichmanMoorman2000} was used to quantify the persistence of similarity between waveform patterns when they were extended by one sample. It was defined as

\begin{equation}
\mathrm{SampEn}(m,r)
=
-\ln\left(\frac{A}{B}\right),
\end{equation}

where $B$ is the number of distinct template pairs of length $m$ satisfying the similarity criterion and $A$ is the number of corresponding pairs that remain similar when extended to length $m+1$. Sample entropy was calculated using $m=2$ and $\tau=1$, with the similarity tolerance set to $r=0.10\sigma$, where $\sigma$ is the population standard deviation of the analyzed segment. Similarity was evaluated using the Chebyshev distance, with self-matches excluded.

Shannon entropy was calculated from the amplitude distribution as

\begin{equation}
H_{\mathrm{S}}
=
-\sum_{k=1}^{K} p_k \ln(p_k),
\end{equation}

where $p_k$ denotes the probability of samples falling within histogram bin $k$. The amplitude range of each segment was divided into $K=50$ equal-width bins.

Natural logarithms were used for all three entropy measures. When sample entropy was undefined because no valid template matches were available, the corresponding feature value was treated as missing and subsequently imputed using statistics estimated from the corresponding cross-validation training partition.

\subsection*{Feature extraction parameter settings}

The peak and burst detection parameters were treated as analysis settings rather than physiological constants. The selected settings were determined during preliminary exploratory development using the TPEHGT recordings and were subsequently fixed for all analyses reported in this study. Because this preliminary parameter selection used the same dataset as the final evaluation and was not nested within the outer cross-validation procedure, the resulting performance estimates are interpreted as exploratory rather than as independent validation of the selected parameter configuration.

The final parameter settings are summarized in Table~\ref{tab:param_selection}. THK and MD were common to both segmentation strategies, whereas the burst-separation threshold BT was set separately because the annotated intervals are variable in duration while the fixed windows have a uniform duration. The sample-entropy parameter $R$ defines the tolerance as $r=R\sigma$, where $\sigma$ is the within-segment population standard deviation.

\begin{table}[!htbp]
\centering
\caption{\textbf{Feature extraction parameter settings.} THK is the multiplier used in the robust peak-detection threshold; MD is the minimum distance between detected peaks; BT is the maximum inter-peak interval used to group adjacent peaks into the same burst-like event; and $R$ is the multiplier applied to the within-segment population standard deviation to define the sample-entropy tolerance $r=R\sigma$.}
\label{tab:param_selection}
\begin{tabular}{lcccc}
\hline
\textbf{Segmentation strategy} &
\textbf{THK} &
\textbf{MD (s)} &
\textbf{BT (s)} &
\textbf{$R$} \\
\hline
Annotated intervals & 2.8 & 0.3 & 1.8 & 0.1 \\
Fixed 3-minute      & 2.8 & 0.3 & 2.0 & 0.1 \\
\hline
\end{tabular}
\end{table}

\subsection*{Classification models}

The extracted feature vectors were evaluated using nine supervised classification models: Quadratic Discriminant Analysis (QDA), Logistic Regression (LR), Support Vector Machine (SVM), Decision Tree (DT), Random Forest (RF), Gradient Boosting (GB), Gaussian Naive Bayes (NB), Multi-Layer Perceptron (MLP), and CatBoost (CB). The model settings used throughout the experiments are summarized in Table~\ref{tab:model_settings}. All stochastic models used a random seed of 42.

\begin{table}[!htbp]
\centering
\caption{\textbf{Classification model settings used in all experiments.} Parameters not explicitly listed retained the defaults of the corresponding software implementation; the complete version-pinned computational environment is provided with the public code repository.}
\label{tab:model_settings}
\begin{tabular}{p{0.13\textwidth} p{0.77\textwidth}}
\hline
\textbf{Model} & \textbf{Settings} \\
\hline
QDA &
Regularization parameter = 0.5 \\

LR &
L-BFGS solver; $C=0.5$; balanced class weights;
maximum iterations = 50,000 \\

SVM &
Linear kernel; $C=2.0$; balanced class weights;
probability estimation enabled \\

DT &
Default estimator settings; random seed = 42 \\

RF &
500 trees; maximum depth = 15; minimum leaf size = 1;
balanced class weights; random seed = 42 \\

GB &
Default estimator settings; random seed = 42 \\

NB &
Gaussian Naive Bayes; default estimator settings \\

MLP &
Hidden layers = 256 and 128 units; L2 penalty = 0.0005;
maximum iterations = 3,000; internal validation-based early
stopping disabled; training-loss convergence patience = 25 iterations;
random seed = 42 \\

CB &
Depth = 6; learning rate = 0.1; logarithmic loss;
random seed = 42 \\
\hline
\end{tabular}
\end{table}

The selected models span linear, probabilistic, kernel-based, tree-based, ensemble, and neural-network approaches. Tree-based ensemble models such as RF, GB, and CB were included because they can represent nonlinear relationships and interactions among the extracted features without strong distributional assumptions. CatBoost was included as an additional gradient-boosting approach for tabular feature-based classification~\cite{Prokhorenkova2018CatBoost}.

For every outer training partition, missing feature values were imputed using means estimated exclusively from that training partition, after which features were standardized using the corresponding training mean and standard deviation. The fitted imputation and standardization transformations were then applied unchanged to the associated validation data. Imputation, standardization, and classification were implemented within a single fold-fitted processing pipeline.

For each segment, the fitted classifier produced an estimated preterm-class probability using its model-specific probability prediction. These outputs were used as classification scores for ranking, recording-level aggregation, and threshold-based classification; they were not interpreted as calibrated individual clinical risk probabilities.

\subsection*{Evaluation metrics}

Model performance was evaluated using 30 randomized repetitions of five-fold stratified cross-validation performed at the recording level. For each repetition, one class label was associated with each of the 26 recordings and the recordings were newly partitioned into five mutually exclusive folds. The resulting recording identifiers were then mapped back to their constituent segments, ensuring that all segments from a given recording remained entirely within either the training or validation partition of each outer fold. Each recording therefore contributed one held-out prediction per repetition.

Within every outer fold, feature imputation and standardization were estimated exclusively from the outer-training data, after which the classifier was fitted to the training segments. Each segment produced an estimated preterm-class output through the classifier's probability prediction. These outputs were treated as classification scores rather than as calibrated individual clinical risk probabilities.

Because the class label was defined at the recording level, segment-level scores were aggregated to obtain one score for each recording. Maximum-score aggregation was used:

\begin{equation}
s_{\mathrm{record}}
=
\max_i s_i,
\label{eq:record_aggregation}
\end{equation}

where $s_i$ denotes the preterm-class score of segment $i$ belonging to the same recording. This aggregation rule was selected to retain the strongest preterm-like segment-level evidence within a recording.

For threshold-dependent classification, a decision threshold was determined separately within each outer-training partition. After model fitting, scores were obtained for the same outer-training segments and aggregated to the recording level using Eq.~\ref{eq:record_aggregation}. Candidate thresholds corresponding to the observed training-record scores were evaluated, and the threshold yielding the highest Matthews correlation coefficient (MCC) on these fitted training-record scores was selected. The resulting threshold was then applied unchanged to the corresponding outer-validation recording scores. Outer-validation labels were not used for model fitting or threshold selection. Because threshold selection used fitted training scores rather than an additional inner out-of-fold procedure, the resulting threshold-dependent performance is interpreted as exploratory.

Preterm delivery was treated as the positive class (label~1). Threshold-dependent performance was summarized using accuracy, F1-score, balanced accuracy, and MCC. Let TP, TN, FP, and FN denote true positives, true negatives, false positives, and false negatives, respectively. Accuracy was defined as

\begin{equation}
\mathrm{Accuracy}
=
\frac{TP+TN}{TP+TN+FP+FN}.
\end{equation}

Precision and recall were defined as

\begin{equation}
\mathrm{Precision}
=
\frac{TP}{TP+FP},
\qquad
\mathrm{Recall}
=
\frac{TP}{TP+FN},
\end{equation}

and the F1-score as

\begin{equation}
\mathrm{F1}
=
2
\frac{\mathrm{Precision}\cdot\mathrm{Recall}}
{\mathrm{Precision}+\mathrm{Recall}}.
\end{equation}

Balanced accuracy was calculated as

\begin{equation}
\mathrm{Balanced\ Accuracy}
=
\frac{1}{2}
\left(
\frac{TP}{TP+FN}
+
\frac{TN}{TN+FP}
\right),
\end{equation}

and MCC as

\begin{equation}
\mathrm{MCC}
=
\frac{TP\cdot TN-FP\cdot FN}
{\sqrt{(TP+FP)(TP+FN)(TN+FP)(TN+FN)}}.
\end{equation}

MCC was used for threshold selection because it incorporates all four entries of the binary confusion matrix and provides a balanced summary of classification performance~\cite{Chicco2020_MCC}.

Threshold-independent discrimination was evaluated using the area under the receiver operating characteristic curve (ROC-AUC) and average precision (AP). Both were calculated directly from the continuous recording-level classification scores and therefore did not depend on the selected binary threshold. AP was calculated as the step-weighted summary of the precision-recall curve rather than as a trapezoidal precision-recall area.

For each repetition, the held-out predictions from the five outer validation folds were combined to form one complete out-of-fold set containing exactly one prediction for each of the 26 recordings. Threshold-dependent metrics were then calculated from the corresponding fold-specific binary predictions, whereas ROC-AUC and AP were calculated from the pooled continuous recording-level scores. This produced one value for each metric per repetition. The values reported in the Results are the means of the 30 repetition-level estimates.

\section*{Results}

Results are reported as means across 30 repetitions of five-fold stratified cross-validation performed at the recording level. Within each repetition, predictions from the five outer validation folds were combined so that each of the 26 recordings contributed exactly one out-of-fold prediction before the recording-level metrics were calculated. The 26 recordings comprised 13 term and 13 preterm recordings originating from 18 pregnancies.

The annotated-interval analysis included 200 dataset-provided segments, comprising 100 contraction intervals and 100 dummy (non-contraction) intervals. The fixed-window analysis included 249 complete, non-overlapping 3-minute segments; 15 recordings contributed ten windows and 11 recordings contributed nine. In all experiments, segment-level classification scores were combined using maximum-score aggregation to obtain one recording-level score before metric calculation. Preterm delivery was treated as the positive class.

\subsection*{Comparison of intrinsic mode functions}

Table~\ref{tab:imf_selection} compares the first four IMFs using fixed 3-minute segmentation under the same feature extraction and recording-level evaluation framework. For each IMF, the classifier with the highest mean balanced accuracy was retained for the comparison. Among the four evaluated modes, IMF1 yielded the highest mean values for all six reported metrics, with RF achieving an accuracy of 0.8308, F1-score of 0.7969, balanced accuracy of 0.8308, MCC of 0.6998, ROC-AUC of 0.8157, and AP of 0.8877. The corresponding mean balanced accuracies for IMF2, IMF3, and IMF4 were 0.7218, 0.5462, and 0.6359, respectively.

IMF1 was therefore carried forward as the empirically selected mode for the subsequent analyses. Because mode and classifier selection were performed using the same dataset on which their cross-validated performance was summarized, this comparison is interpreted as an exploratory within-dataset selection rather than an independent validation of IMF1.

\begin{table}[!htbp]
\centering
\caption{\textbf{Recording-level comparison of the first four IMFs using fixed 3-minute segmentation.} For each IMF, the classifier with the highest mean balanced accuracy was retained. Values are means across 30 complete repetition-level out-of-fold evaluations. Preterm delivery was treated as the positive class.}
\label{tab:imf_selection}
\resizebox{\textwidth}{!}{%
\setlength{\tabcolsep}{4pt}
\begin{tabular}{lccccccc}
\hline
\textbf{IMF} & \textbf{Best model} & \textbf{Accuracy} & \textbf{F1} & \textbf{Balanced Acc.} & \textbf{MCC} & \textbf{ROC-AUC} & \textbf{AP} \\
\hline
\textbf{IMF1} & RF &
$\mathbf{0.8308}$ &
$\mathbf{0.7969}$ &
$\mathbf{0.8308}$ &
$\mathbf{0.6998}$ &
$\mathbf{0.8157}$ &
$\mathbf{0.8877}$ \\

IMF2 & RF &
$0.7218$ &
$0.6520$ &
$0.7218$ &
$0.4849$ &
$0.7403$ &
$0.7953$ \\

IMF3 & RF &
$0.5462$ &
$0.3434$ &
$0.5462$ &
$0.1119$ &
$0.6440$ &
$0.6342$ \\

IMF4 & GB &
$0.6359$ &
$0.6347$ &
$0.6359$ &
$0.2756$ &
$0.6744$ &
$0.6839$ \\
\hline
\end{tabular}%
}
\end{table}
\FloatBarrier

The representative decomposition in Fig.~\ref{fig:imf_acf_psd} shows that IMF1 contains faster local variation than the subsequent modes. The class-averaged PSDs in Fig.~\ref{fig:psd} also show that IMF1 is not confined to a fixed narrow frequency band. Preterm recordings had higher mean IMF1 power mainly below approximately 0.7~Hz, whereas term recordings had higher power over parts of the middle of the displayed range. These curves are descriptive and do not establish a single physiological frequency interval for IMF1.

\subsection*{Annotated-interval classification results}

Table~\ref{tab:contract_ehg} summarizes the recording-level classification results obtained from IMF1 features extracted from the dataset-provided annotated intervals. RF yielded the highest mean values across all six reported metrics, with an accuracy of 0.7821, F1-score of 0.7793, balanced accuracy of 0.7821, MCC of 0.5687, ROC-AUC of 0.8023, and AP of 0.8817. GB and CB also showed comparatively high mean performance among the evaluated classifiers.

\begin{table}[!htbp]
\centering
\caption{\textbf{Recording-level classification performance for annotated-interval segmentation using IMF1 features.} Values are means across 30 complete repetition-level out-of-fold evaluations. Preterm delivery was treated as the positive class.}
\label{tab:contract_ehg}
\resizebox{\textwidth}{!}{%
\setlength{\tabcolsep}{4pt}
\begin{tabular}{lcccccc}
\hline
\textbf{Model} & \textbf{Accuracy} & \textbf{F1} & \textbf{Balanced Acc.} & \textbf{MCC} & \textbf{ROC-AUC} & \textbf{AP} \\
\hline
QDA & $0.5538$ & $0.4337$ & $0.5538$ & $0.1184$ & $0.5866$ & $0.5680$ \\
LR  & $0.6577$ & $0.6188$ & $0.6577$ & $0.3273$ & $0.6907$ & $0.7370$ \\
SVM & $0.5359$ & $0.5926$ & $0.5359$ & $0.0746$ & $0.5836$ & $0.6336$ \\
DT  & $0.5513$ & $0.6479$ & $0.5513$ & $0.1214$ & $0.5513$ & $0.5290$ \\
RF  & $\mathbf{0.7821}$ & $\mathbf{0.7793}$ & $\mathbf{0.7821}$ & $\mathbf{0.5687}$ & $\mathbf{0.8023}$ & $\mathbf{0.8817}$ \\
GB  & $0.7590$ & $0.7572$ & $0.7590$ & $0.5221$ & $0.7784$ & $0.8450$ \\
NB  & $0.6641$ & $0.6200$ & $0.6641$ & $0.3381$ & $0.6448$ & $0.6370$ \\
MLP & $0.6269$ & $0.6393$ & $0.6269$ & $0.2556$ & $0.6483$ & $0.7019$ \\
CB  & $0.7538$ & $0.7222$ & $0.7538$ & $0.5265$ & $0.7921$ & $0.8601$ \\
\hline
\end{tabular}%
}
\end{table}
\FloatBarrier

\subsection*{Fixed 3-minute classification results}

Table~\ref{tab:fixed_ehg} summarizes the recording-level classification results obtained from fixed 3-minute IMF1 segments. RF yielded the highest mean values across all six reported metrics, achieving an accuracy of 0.8308, F1-score of 0.7969, balanced accuracy of 0.8308, MCC of 0.6998, ROC-AUC of 0.8157, and AP of 0.8877. Among the configurations evaluated in this study, fixed 3-minute IMF1 features with RF yielded the highest observed mean performance across all six metrics.

\begin{table}[!htbp]
\centering
\caption{\textbf{Recording-level performance for fixed 3-minute segmentation using EHG-only IMF1 features.} Values are means across 30 complete repetition-level out-of-fold evaluations. Preterm delivery was treated as the positive class.}
\label{tab:fixed_ehg}
\resizebox{\textwidth}{!}{%
\setlength{\tabcolsep}{4pt}
\begin{tabular}{lcccccc}
\hline
\textbf{Model} & \textbf{Accuracy} & \textbf{F1} & \textbf{Balanced Acc.} & \textbf{MCC} & \textbf{ROC-AUC} & \textbf{AP} \\
\hline
QDA & $0.5513$ & $0.5547$ & $0.5513$ & $0.1034$ & $0.5193$ & $0.5498$ \\
LR  & $0.6051$ & $0.5886$ & $0.6051$ & $0.2137$ & $0.6402$ & $0.6756$ \\
SVM & $0.5603$ & $0.5723$ & $0.5603$ & $0.1221$ & $0.5584$ & $0.6114$ \\
DT  & $0.5205$ & $0.6486$ & $0.5205$ & $0.0459$ & $0.5205$ & $0.5122$ \\
RF  & $\mathbf{0.8308}$ & $\mathbf{0.7969}$ & $\mathbf{0.8308}$ & $\mathbf{0.6998}$ & $\mathbf{0.8157}$ & $\mathbf{0.8877}$ \\
GB  & $0.7692$ & $0.7486$ & $0.7692$ & $0.5497$ & $0.7558$ & $0.7961$ \\
NB  & $0.6154$ & $0.5661$ & $0.6154$ & $0.2387$ & $0.6061$ & $0.5955$ \\
MLP & $0.6205$ & $0.5950$ & $0.6205$ & $0.2439$ & $0.6114$ & $0.6533$ \\
CB  & $0.8013$ & $0.7586$ & $0.8013$ & $0.6445$ & $0.8034$ & $0.8773$ \\
\hline
\end{tabular}%
}
\end{table}
\FloatBarrier
 
\begin{figure}[!htbp]
\centering
\includegraphics[width=\textwidth]{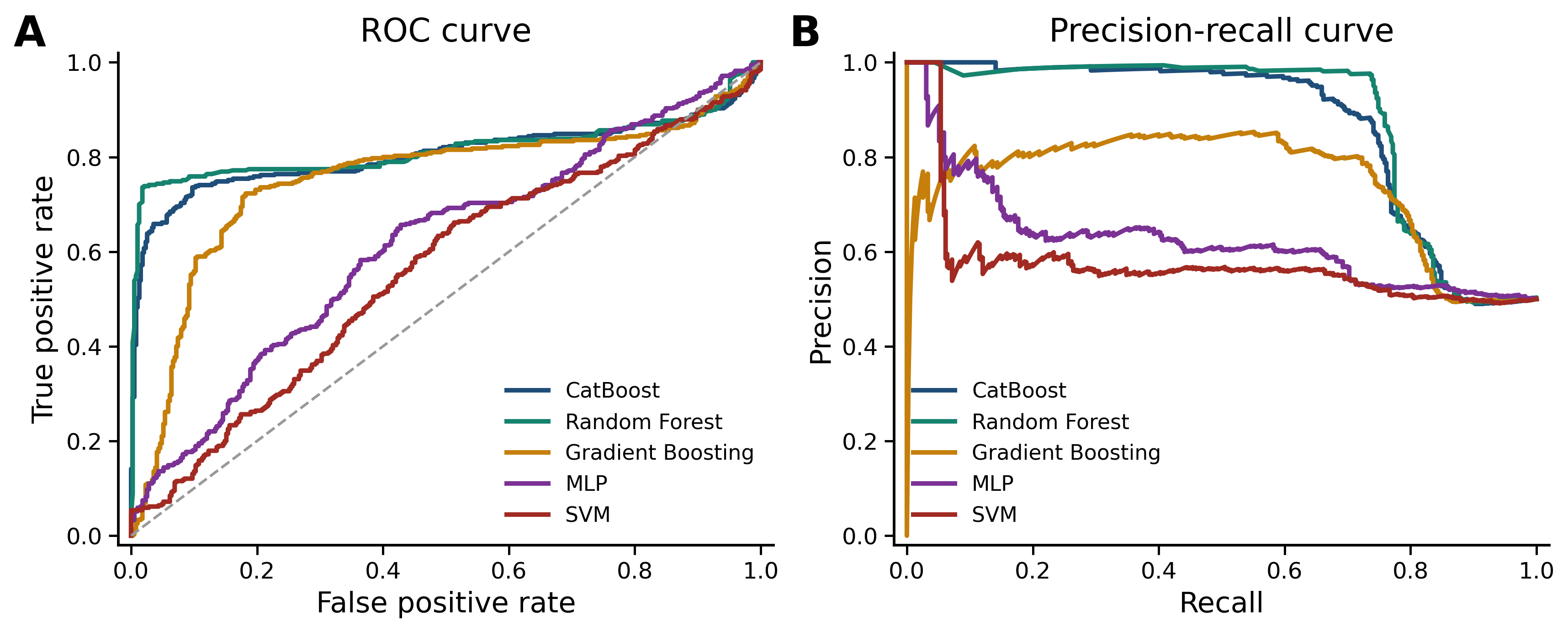}
\caption{\textbf{Recording-level ROC and precision--recall curves for fixed 3-minute IMF1 features.} (A)~Receiver operating characteristic curves. (B)~Precision-recall curves. Curves were generated by pooling the repeated out-of-fold recording-level classification scores obtained across the 30 cross-validation repetitions. Within each repetition, each of the 26 recordings contributed exactly one out-of-fold score after maximum-score aggregation of its segment-level outputs. The pooled curves therefore contain repeated out-of-fold predictions of the same 26 recordings across different data partitions and should not be interpreted as representing 780 independent observations. The ROC-AUC and AP values associated with these pooled curves are descriptive quantities and are distinct from the mean repetition-level ROC-AUC and AP values reported in Table~\ref{tab:fixed_ehg}.}
\label{fig:roc_pr}
\end{figure}

\subsection*{Comparison with filtered time-domain features}

To isolate the effect of the EMD-derived representation, the same feature definitions, segmentation, classification models, recording-level aggregation, and evaluation procedure were applied directly to the dataset-filtered time-domain EHG signals. Thus, the IMF1 and filtered-signal analyses differed in the input signal representation while retaining the same downstream feature and classification framework.

Under fixed 3-minute segmentation, RF using IMF1 features yielded a mean accuracy of 0.8308 compared with 0.7846 for the matched filtered-signal representation. F1-score increased from 0.7376 to 0.7969, balanced accuracy from 0.7846 to 0.8308, and MCC from 0.6114 to 0.6998. The corresponding threshold-independent metrics were also numerically higher for IMF1, with ROC-AUC increasing from 0.8073 to 0.8157 and AP from 0.8773 to 0.8877 (Table~\ref{tab:fixed_raw}).

Therefore, IMF1 yielded numerically higher mean performance across all six reported metrics in the matched fixed-window RF comparison. The differences in ROC-AUC and AP were modest, however, and the representation comparison is interpreted descriptively because no independent test set or paired inferential comparison was used.

\begin{table}[!htbp]
\centering
\caption{\textbf{Recording-level classification performance for fixed 3-minute segmentation using matched filtered time-domain EHG features.} Values are means across 30 complete repetition-level out-of-fold evaluations. Preterm delivery was treated as the positive class.}
\label{tab:fixed_raw}
\resizebox{\textwidth}{!}{%
\setlength{\tabcolsep}{4pt}
\begin{tabular}{lcccccc}
\hline
\textbf{Model} & \textbf{Accuracy} & \textbf{F1} & \textbf{Balanced Acc.} & \textbf{MCC} & \textbf{ROC-AUC} & \textbf{AP} \\
\hline
QDA & $0.6487$ & $0.6752$ & $0.6487$ & $0.3029$ & $0.6286$ & $0.6198$ \\
LR  & $0.6590$ & $0.6740$ & $0.6590$ & $0.3215$ & $0.7231$ & $0.6965$ \\
SVM & $0.6756$ & $0.6978$ & $0.6756$ & $0.3579$ & $0.6974$ & $0.6967$ \\
DT  & $0.5308$ & $0.6541$ & $0.5308$ & $0.0825$ & $0.5308$ & $0.5172$ \\
RF  & $\mathbf{0.7846}$ & $\mathbf{0.7376}$ & $\mathbf{0.7846}$ & $\mathbf{0.6114}$ & $\mathbf{0.8073}$ & $\mathbf{0.8773}$ \\
GB  & $0.7551$ & $0.7205$ & $0.7551$ & $0.5284$ & $0.7624$ & $0.7871$ \\
NB  & $0.6590$ & $0.6304$ & $0.6590$ & $0.3260$ & $0.6733$ & $0.7120$ \\
MLP & $0.6705$ & $0.6887$ & $0.6705$ & $0.3446$ & $0.7114$ & $0.7557$ \\
CB  & $0.7513$ & $0.6902$ & $0.7513$ & $0.5473$ & $0.8051$ & $0.8717$ \\
\hline
\end{tabular}%
}
\end{table}
\FloatBarrier

\subsection*{Overall comparison}

Table~\ref{tab:best_overall_comparison} summarizes the four principal segmentation and signal-representation configurations. RF had the highest mean balanced accuracy in each of the four settings. Within both segmentation strategies, the IMF1 representation yielded numerically higher mean accuracy, F1-score, balanced accuracy, MCC, ROC-AUC, and AP than the matched filtered time-domain representation.

Among the four configurations, fixed 3-minute IMF1 features with RF yielded the highest observed mean value for all six reported metrics. This configuration is therefore treated as the empirically selected best-observed setting in the present dataset. Because the classifier, signal representation, and segmentation comparisons were performed using the same repeated cross-validation results used for performance reporting, the comparison is exploratory rather than an independent validation of the selected configuration.

\begin{table}[!htbp]
\centering
\caption{\textbf{Exploratory best-observed recording-level performance across segmentation strategies and signal representations.} For each configuration, the classifier with the highest mean balanced accuracy was retained. Values are means across 30 complete repetition-level out-of-fold evaluations. Model, representation, and segmentation comparisons were performed within the same dataset and should therefore be interpreted as exploratory. Preterm delivery was treated as the positive class.}
\label{tab:best_overall_comparison}
\resizebox{\textwidth}{!}{%
\setlength{\tabcolsep}{3pt}
\begin{tabular}{llccccccc}
\hline
\textbf{Segmentation} &
\textbf{Feature source} &
\textbf{Best model} &
\textbf{Accuracy} &
\textbf{F1} &
\textbf{Balanced Acc.} &
\textbf{MCC} &
\textbf{ROC-AUC} &
\textbf{AP} \\
\hline

Annotated intervals &
Filtered time-domain &
RF &
$0.7167$ &
$0.7032$ &
$0.7167$ &
$0.4382$ &
$0.7719$ &
$0.8411$ \\

Annotated intervals &
IMF1 &
RF &
$0.7821$ &
$0.7793$ &
$0.7821$ &
$0.5687$ &
$0.8023$ &
$0.8817$ \\

Fixed 3-minute &
Filtered time-domain &
RF &
$0.7846$ &
$0.7376$ &
$0.7846$ &
$0.6114$ &
$0.8073$ &
$0.8773$ \\

\textbf{Fixed 3-minute} &
\textbf{IMF1} &
\textbf{RF} &
$\mathbf{0.8308}$ &
$\mathbf{0.7969}$ &
$\mathbf{0.8308}$ &
$\mathbf{0.6998}$ &
$\mathbf{0.8157}$ &
$\mathbf{0.8877}$ \\

\hline
\end{tabular}%
}
\end{table}
\FloatBarrier

\begin{figure}[!htbp]
\centering
\includegraphics[width=\textwidth]{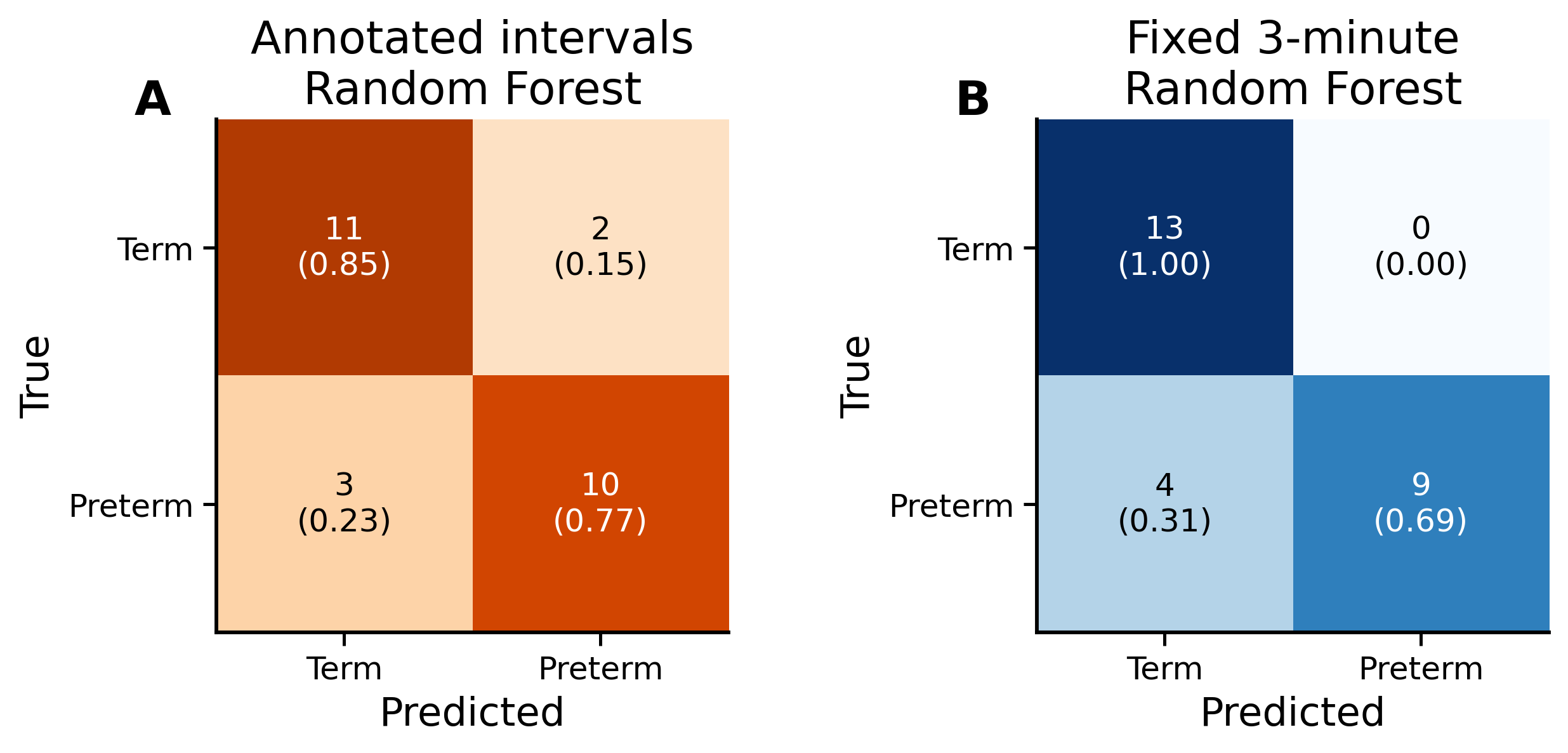}
\caption{\textbf{Majority-vote consensus recording-level confusion matrices for selected RF configurations.} Rows correspond to true class labels and columns correspond to predicted labels. Each displayed prediction represents the majority-vote consensus across the 30 out-of-fold binary predictions available for that recording. Within each outer fold, segment-level classification scores were combined using maximum-score aggregation, and the MCC-optimized threshold selected from the corresponding fitted training-record scores was applied unchanged to the validation recording scores. Values in parentheses show the row-normalized fractions. These matrices provide descriptive consensus summaries across repeated cross-validation partitions and do not represent the repetition-level confusion matrices used to calculate the mean performance metrics in Table~\ref{tab:best_overall_comparison}.}
\label{fig:confusion}
\end{figure}

To complement the repetition-level performance summaries in
Table~\ref{tab:best_overall_comparison}, Fig.~\ref{fig:confusion} shows a
consensus recording-level classification summary for the selected RF
configurations. Each recording had one out-of-fold binary prediction in
each of the 30 cross-validation repetitions, and the displayed class was
determined by majority vote across these 30 predictions.

Under this consensus summary, the fixed 3-minute IMF1 configuration
correctly classified all 13 term recordings and 9 of 13 preterm recordings,
whereas the annotated-interval IMF1 configuration correctly classified
11 of 13 term recordings and 10 of 13 preterm recordings. The fixed-window
configuration therefore produced four consensus errors compared with five
for the annotated-interval configuration and produced no false-positive
preterm classifications among the term recordings. In contrast, the
annotated-interval configuration correctly identified one additional
preterm recording.

The consensus matrices are descriptive summaries of prediction stability
across repeated cross-validation partitions and are distinct from the
repetition-level confusion matrices used to calculate the mean performance
metrics reported in Table~\ref{tab:best_overall_comparison}.

\subsection*{Feature behavior at the recording level}

\begin{figure}[!htbp]
\centering
\includegraphics[width=\textwidth]{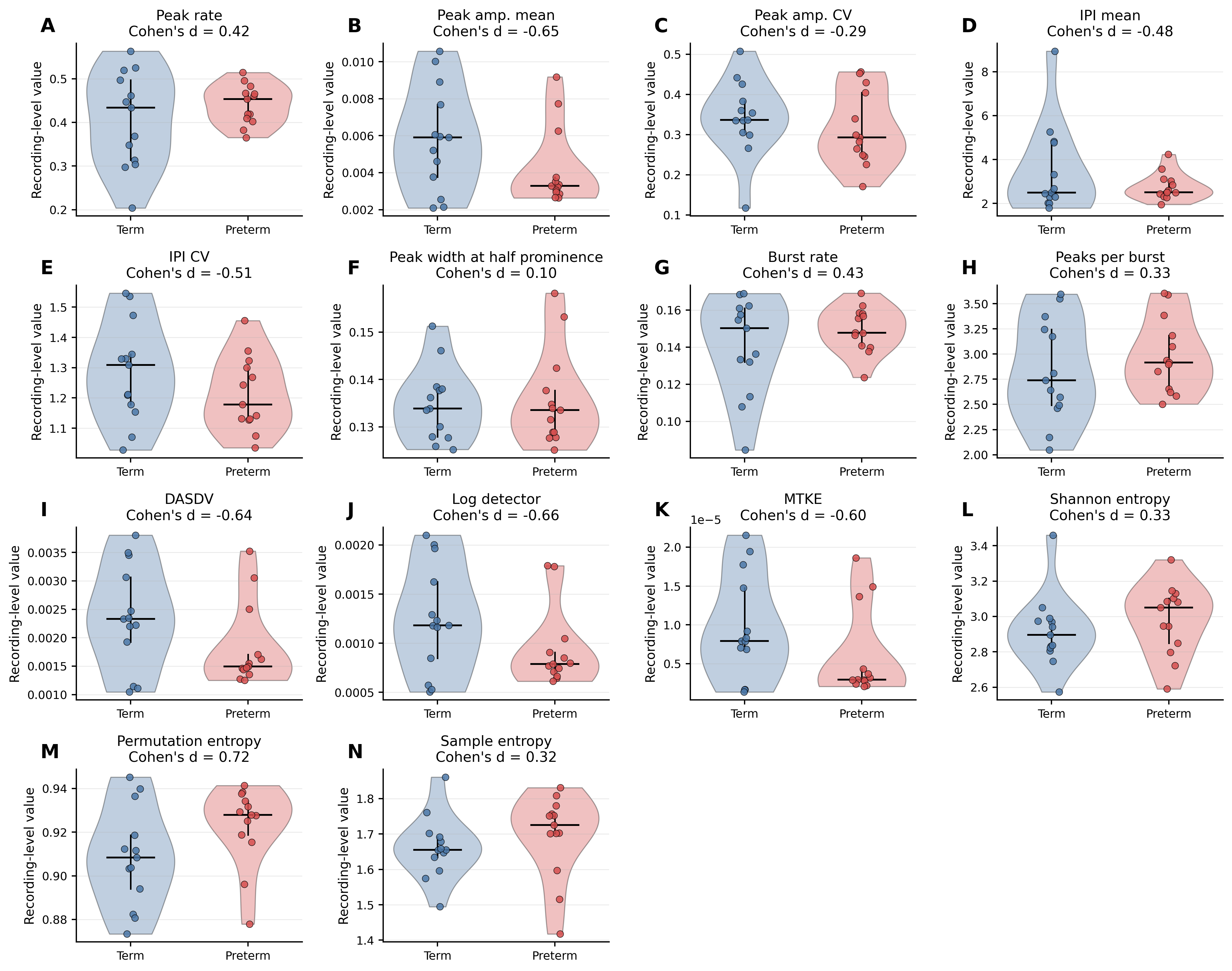}
\caption{\textbf{Recording-level distributions of the 14 IMF1 feature types for term and preterm recordings in the fixed 3-minute analysis.} Each feature was extracted independently from the three EHG channels, averaged across channels for visualization, and subsequently averaged across fixed 3-minute segments within each recording. The classifiers used the full 42-dimensional channel-specific feature vector, whereas this figure provides a compact feature-type-level summary for signal-level interpretation. Violin shapes show the distribution within each class, horizontal black lines indicate the median, and individual recording values are overlaid as points. Cohen's $d$ denotes the standardized difference between preterm and term recording-level feature means, with positive values indicating higher values in preterm recordings. The effect sizes are descriptive and were not accompanied by feature-level hypothesis tests or multiplicity correction.}
\label{fig:feature_dist}
\end{figure}

To examine the signal characteristics underlying the selected fixed 3-minute IMF1 configuration, recording-level feature distributions were summarized by feature type in Fig.~\ref{fig:feature_dist}. Although the classifiers used the full 42-dimensional channel-specific feature vector, each of the fourteen feature types was averaged across the three EHG channels for visualization, and segment-level values were subsequently averaged within each recording. This produced one recording-level value per feature type for each of the 26 recordings.

The distributions show class-dependent differences across multiple feature families rather than a uniform increase or decrease in all IMF1 descriptors. Features with positive Cohen's $d$ values, indicating higher recording-level values in the preterm group, included peak rate, peak width at half prominence, burst rate, mean peaks per burst, Shannon entropy, permutation entropy, and sample entropy. Permutation entropy showed the largest positive effect size, whereas peak width at half prominence showed only a small difference. Features with negative Cohen's $d$ values included mean peak amplitude, peak-amplitude coefficient of variation, mean inter-peak interval, inter-peak-interval coefficient of variation, DASDV, log detector, and MTKE.

Taken together, preterm recordings showed more frequent and more densely grouped peak-like IMF1 events, with shorter and less variable inter-peak spacing and smaller, less variable peak amplitudes. DASDV, log detector, and MTKE were also lower, whereas Shannon, permutation, and sample entropy were higher. The combined pattern therefore reflects differences in detected event organization, waveform magnitude and temporal energy, and signal complexity rather than a single common change in overall variability.

The largest descriptive effect sizes were observed for permutation entropy, log detector, mean peak amplitude, DASDV, and MTKE. These distributions describe recording-level directions of class separation but do not establish any individual descriptor as a standalone physiological biomarker. No feature-level hypothesis tests were performed and no multiplicity-adjusted statistical significance is inferred from the Cohen's $d$ values. Moreover, because the public dataset does not provide a reproducible mother-to-recording linkage, possible dependence between recordings originating from the same pregnancy is not represented in these effect-size calculations. The distributions should therefore be interpreted as descriptive signal-level complements to the multivariate model analysis.

Cohen's $d$ was used to quantify the standardized difference between the preterm and term recording-level feature means:

\begin{equation}
d =
\frac{\mu_{\mathrm{preterm}}-\mu_{\mathrm{term}}}
{s_{\mathrm{pooled}}},
\end{equation}

where $\mu_{\mathrm{preterm}}$ and $\mu_{\mathrm{term}}$ denote the preterm and term recording-level feature means, respectively, and

\begin{equation}
s_{\mathrm{pooled}}
=
\sqrt{
\frac{
(n_{\mathrm{preterm}}-1)s_{\mathrm{preterm}}^2
+
(n_{\mathrm{term}}-1)s_{\mathrm{term}}^2
}{
n_{\mathrm{preterm}}+n_{\mathrm{term}}-2
}
}.
\end{equation}

With this definition, positive Cohen's $d$ values indicate higher recording-level feature values in preterm recordings, whereas negative values indicate higher values in term recordings.

\subsection*{Feature importance analysis}

Grouped permutation feature importance was used to examine the dependence of the selected IMF1 RF models on three predefined feature families: burst and peak features, temporal-energy features, and entropy features. Each group contained the corresponding channel-specific features from all three EHG channels, thereby preserving the same 42-dimensional feature representation used by the classifiers.

Within each outer validation fold, all features belonging to a given family were jointly permuted across the validation segment rows using the same row permutation for every feature in that family. This preserved the relationships among features within the permuted family while disrupting their association with the original validation segments. One permutation was performed for each feature family in each outer fold and cross-validation repetition, using a base random seed of 42.

After permutation, segment-level classification scores were combined using the same maximum-score recording-level aggregation as in the primary analysis. Within each of the 30 cross-validation repetitions, the baseline and permuted recording-level scores from the five outer validation folds were pooled to form complete 26-record out-of-fold sets. Average precision (AP) was then calculated once for the baseline scores and once for each permuted feature family, and importance was defined as

\begin{equation}
I_g =
\mathrm{AP}_{\mathrm{baseline}}
-
\mathrm{AP}_{\mathrm{permuted},g},
\end{equation}

where $g$ denotes the feature family. The reported importance for each feature family is the mean of the 30 repetition-level AP decreases. Larger positive values indicate greater dependence of the fitted RF model on the corresponding feature family, whereas values near zero indicate little change in recording-level AP after permutation.

\begin{figure}[!htbp]
\centering
\includegraphics[width=0.80\textwidth]{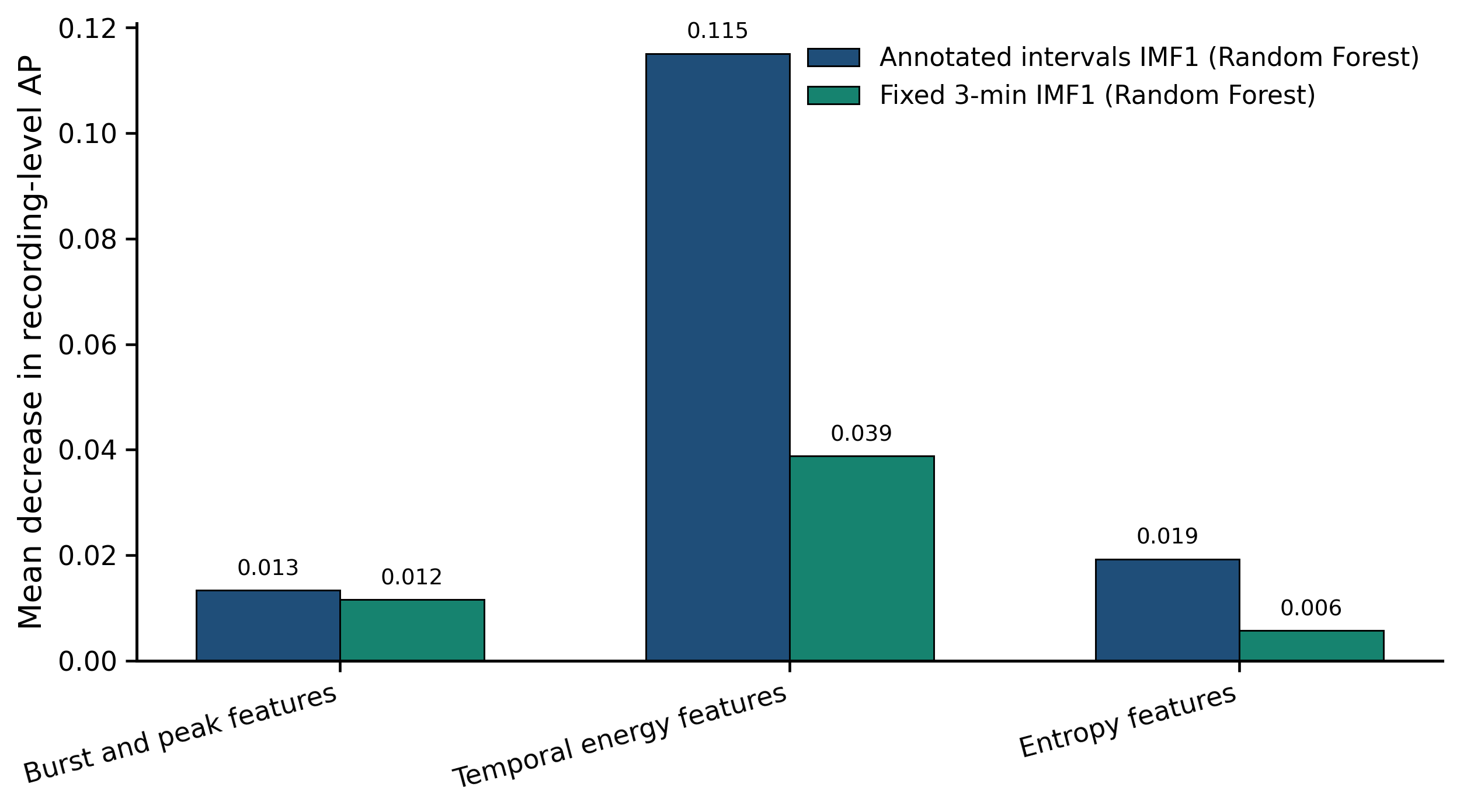}
\caption{\textbf{Grouped permutation feature importance for the annotated-interval and fixed 3-minute IMF1 RF models.} Importance is shown as the mean decrease in recording-level average precision (AP) after jointly permuting all features within each feature family across validation segment rows. Within each of the 30 cross-validation repetitions, baseline and permuted recording-level scores from the five outer validation folds were pooled to form complete 26-record out-of-fold sets before AP was calculated. One permutation was performed per feature family per outer fold in each repetition. Larger positive decreases indicate greater model dependence on the corresponding feature family, whereas values near zero indicate little change in recording-level AP after permutation.}
\label{fig:grouped_feature_importance}
\end{figure}

Grouped importance was calculated for the RF models in both the annotated-interval IMF1 and fixed 3-minute IMF1 settings. As shown in Fig.~\ref{fig:grouped_feature_importance}, temporal-energy features produced the largest mean decrease in recording-level AP under both segmentation strategies. In the annotated-interval IMF1 model, permutation of the temporal-energy features decreased AP by 0.1151 on average, compared with 0.0193 for entropy features and 0.0133 for burst and peak features. In the fixed 3-minute IMF1 model, the corresponding mean decreases were 0.0388, 0.0057, and 0.0116, respectively.

Thus, temporal-energy descriptors produced the largest mean AP decrease in both selected RF models. Burst and peak features showed smaller positive mean importance in both settings, while the entropy group also showed a positive but smaller model-level contribution than the temporal-energy group. These model-level results should be distinguished from the descriptive feature distributions in Fig.~\ref{fig:feature_dist}: the consistent increase in entropy measures observed in preterm recordings does not imply that entropy provides the largest unique contribution to the RF classifiers. Because information may also be shared among correlated feature families, grouped permutation importance is interpreted as a descriptive measure of model dependence rather than as evidence of an independent physiological mechanism.

\section*{Discussion}

\subsection*{Main findings}
This study evaluated a signal-adaptive and interpretable EHG framework under recording-level validation designed to prevent direct segment leakage across training and validation folds. IMF1 yielded the highest mean performance among the first four evaluated EMD modes, fixed 3-minute windows yielded higher mean performance than the dataset-provided annotated intervals, and the IMF1 representation produced higher mean threshold-dependent performance than matched filtered time-domain features under fixed-window segmentation. 

Together, these findings support further evaluation of EMD as a signal-adaptive representation for recording-level term-versus-preterm EHG classification. All segments from a given recording remained within the same outer cross-validation fold, preventing direct sharing of recording-specific segments between training and validation data. However, pregnancy-wise independence could not be established from the identifiers provided with the public dataset. The findings should therefore be interpreted as exploratory recording-level evidence rather than independent pregnancy-level validation or calibrated clinical risk prediction.

\subsection*{Interpretation of IMF1-derived feature patterns}
A central feature of the preterm-associated pattern was the coexistence of reduced waveform magnitude and event-level dispersion with increased entropy. Lower DASDV, log detector, and MTKE indicate smaller short-lag changes, lower typical IMF1 magnitude, and lower mean temporal energy, while lower peak-amplitude and inter-peak-interval coefficients of variation indicate more homogeneous detected event size and spacing. These observations are not inconsistent with higher entropy because the measures describe different signal properties: Shannon entropy summarizes amplitude-bin occupancy, permutation entropy the diversity of short ordinal patterns, and sample entropy the continuation of local waveform similarity. Combined with the higher peak rate, burst rate, and mean number of peaks per burst, the pattern is consistent with more frequent, smaller, and relatively stereotyped peak-like events embedded within a more complex whole-segment waveform organization.

One possible explanation is alternation between burst-rich and relatively quiescent signal states, with reduced within-state dispersion but greater diversity of state occupancy, transitions, or local waveform morphology. This interpretation therefore represents a testable signal-level hypothesis rather than a demonstrated physiological mechanism.

The weak grouped entropy importance further indicates that the entropy direction was descriptively informative but did not provide strong unique predictive information in the selected RF models; temporal-energy features remained the stronger model-level contribution.

\subsection*{Contribution of EMD and selection of IMF1}
This study builds on earlier EMD-based uterine EMG/EHG research and extends it in four important ways. Previous studies used EMD-derived instantaneous amplitude, instantaneous frequency, entropy ratios, and time-domain descriptors for PTB-related classification tasks~\cite{Ren2015EMD,janjarasjitt2021preterm,janjarasjitt2022comparison}. The present study advances this line of work by evaluating the first four IMFs under a common feature extraction and classification framework, comparing the selected IMF representation with matched filtered time-domain descriptors, comparing annotated-interval and fixed-window segmentation, and reporting performance at the recording level using grouped cross-validation without oversampling.

More recent work has investigated multivariate EMD (MEMD) for multichannel EHG analysis, allowing jointly decomposed channels to maintain greater cross-channel mode consistency \cite{Cui2025MEMD}. The present study instead evaluates standard channel-wise EMD as a simpler reproducible representation and focuses on matched representation, segmentation, and recording-level comparisons. MEMD therefore represents an important future comparator rather than an analysis performed in the present study.

This contribution addresses an important methodological gap in EHG-based PTB prediction. Reported performance in previous studies varies widely and can be strongly influenced by feature representation, segmentation strategy, data partitioning, and whether segments from the same recording are treated as independent samples. By assigning all segments from the same recording to the same validation fold and aggregating segment-level classification scores to one recording-level score, the present study aligns evaluation with the recording-level classification target and avoids treating multiple segments from the same recording as independent validation cases. This makes the reported estimates more appropriate for the study design than segment-level performance measures alone.

IMF1 provided stronger classification performance than IMF2-IMF4, indicating that the first extracted mode preserved the most useful information for the selected peak, burst, temporal energy, and entropy descriptors. The PSD in Fig.~\ref{fig:psd} shows that class differences within IMF1 were distributed across the displayed frequency range and were most apparent in the lower-frequency portion of IMF1. This supports the role of IMF1 as an adaptive EHG representation that can capture discriminative oscillatory structure without requiring a predefined uterine frequency band. This interpretation is consistent with recent EHG work emphasizing the value of interpretable low-frequency EHG information for premature birth prediction~\cite{Pirnar2024}.

Relative to the filtered time-domain baseline, IMF1 produced numerically higher mean accuracy, F1-score, balanced accuracy, MCC, ROC-AUC, and AP. Because the comparison was performed within the same dataset without an independent test cohort or paired inferential analysis, these differences are interpreted descriptively rather than as evidence that EMD is universally superior to the filtered time-domain representation.

\subsection*{Effect of segmentation strategy}
The comparison between dataset-provided annotated intervals and fixed-window segmentation addresses whether useful recording-level information is restricted to manually identified intervals. In this dataset, fixed 3-minute windows yielded higher mean recording-level performance than the annotated-interval strategy, indicating that useful classification information was not confined to annotated contractions.

This finding is consistent with previous TPEHGT work showing that dummy (non-contraction) intervals can also contain discriminative information \cite{Jager2018_TPEHGT}. However, the present comparison does not isolate a single physiological cause for the improvement. Fixed windows also provide uniform segment duration, a larger and more regular set of segments, different opportunities for maximum-score aggregation, and potentially different exposure to transient artifacts. The result should therefore be interpreted as support for annotation-independent fixed-window analysis rather than proof that inter-contraction physiology alone caused the performance difference.

The fixed-window strategy also offers practical advantages for future monitoring systems. It provides a uniform segment length, increases the amount of usable training data, and avoids dependence on manual contraction annotation. These properties make fixed-window analysis well suited to recording-level risk assessment, where the objective is to summarize the electrophysiological state of the uterus across a continuous recording rather than to classify isolated contractions alone. The present findings therefore support fixed-window IMF1 analysis as the preferred segmentation strategy within the proposed framework.

\subsection*{Methodological implications for future PTB risk assessment}
The immediate implication is methodological. Non-overlapping fixed windows provide a standardized, annotation-independent route from continuous EHG recordings to recording-level classification, while the matched filtered-signal comparison helps distinguish the effect of signal representation from differences introduced by the feature set or classifier. The compact feature families also provide a testable multi-scale hypothesis: relatively homogeneous low-amplitude events may occur within a more complex whole-segment organization. These properties make the framework suitable for evaluation in larger linked cohorts; they do not yet establish a calibrated PTB risk score, a clinical decision threshold, or incremental value over obstetric assessment.

\subsection*{Model behavior}
Tree-based ensemble models, particularly RF, GB, and CB, generally outperformed the linear and probabilistic models. This model ranking is consistent with nonlinear effects or interactions between burst, temporal-energy, and entropy descriptors in their relationship with PTB class labels. The weak performance of a single decision tree, compared with the ensembles, also suggests that averaging or boosting helped stabilize predictions in the 42-dimensional feature space.

This model behavior supports the use of conventional feature-based ensemble models as a practical middle ground between simple linear classifiers and fully latent end-to-end deep-learning approaches. Because the models operate on explicit, physiologically understandable feature families, their behavior can be examined post hoc through grouped permutation importance, providing a more auditable prediction framework than an entirely latent representation. RF provided the strongest overall balance between classification performance and this form of model auditability in the proposed framework. The grouped reliance analysis therefore provides model-level evidence of which feature families contribute to prediction, rather than a physiological causal explanation or an explanation of individual patient-level decisions. This emphasis on retaining traceable signal features and examining model reliance aligns with recent calls for more explainable EHG-based prediction methods rather than increasingly complex black-box models~\cite{Pirnar2024,Goldsztejn2023_EHGDeepLearning,Fischer2023_EHGDeepLearning}.

\subsection*{Methodological strengths}
A principal methodological strength is the matched comparison of IMF1 and the filtered time-domain signal under identical segmentation, feature definitions, classifiers, aggregation, and recording-level evaluation. This isolates the effect of representation more directly than comparisons across pipelines with different features or validation procedures. In addition, all segments from one recording remained within the same outer fold, predictions were aggregated to the recording level, and no oversampling was used. These choices directly address avoidable segment leakage and synthetic-sample leakage. A further analytical strength is the joint use of recording-level effect directions and grouped permutation importance, which provides complementary views of the feature space: effect directions characterize how signal properties differ between the classes, while grouped permutation importance assesses the extent to which the trained models rely on each feature family. 
 
\subsection*{Limitations}

This exploratory analysis used 26 recordings originating from 18 pregnancies and no external cohort. Recording-grouped cross-validation prevented fixed windows or annotated intervals from the same recording from crossing the outer split, but the public release does not provide a defensible recording-to-pregnancy key; residual dependence between multiple recordings from the same pregnancy therefore cannot be excluded. Repeated cross-validation assessed the stability of performance across different data partitions, but it does not overcome the limited number of independent pregnancies in the dataset. 

Model scores were not calibrated or evaluated for clinical utility or incremental value over obstetric predictors. The feature distributions and grouped importance analyses are descriptive and do not establish physiological mechanisms or standalone biomarkers. These limitations position the findings as methodological evidence and hypothesis generation for larger cohorts with explicit participant linkage.

\subsection*{Future validation}
Future work should evaluate the prespecified pipeline in larger external cohorts to establish robustness and generalizability, with explicit participant linkage to enable pregnancy-wise independent validation. The larger sample size would also allow performance to be assessed within defined gestational-age windows and prediction horizons. Sensitivity analyses should also examine the influence of proximity to delivery and signal quality. Only after this stage should calibration, clinical utility, and incremental value beyond gestational age, obstetric history, cervical length, or fetal fibronectin be assessed.

\section*{Conclusion}

This study evaluated an interpretable EMD-based framework for recording-level term-versus-preterm classification from non-invasive EHG recordings. Using the public TPEHGT dataset, the first four IMFs were evaluated within a common 42-feature classification framework, with IMF1 yielding the highest mean performance among the evaluated modes. Fixed non-overlapping 3-minute windows also yielded higher mean recording-level performance than the dataset-provided annotated intervals, supporting annotation-independent segmentation as a practical representation of both contraction and inter-contraction EHG activity.

The best-observed configuration, a Random Forest using fixed-window IMF1 features, achieved mean accuracy of $0.8308$, F1-score of $0.7969$, balanced accuracy of $0.8308$, MCC of $0.6998$, ROC-AUC of $0.8157$, and average precision of $0.8877$. Relative to the matched filtered time-domain representation, IMF1 produced numerically higher mean values across all six reported metrics. Recording-level feature distributions showed a preterm-associated pattern of more frequent, smaller, and more regularly spaced peak-like events, lower temporal-energy measures, and higher entropy, while temporal-energy descriptors produced the largest mean grouped permutation importance.

The principal contribution of this work is therefore a transparent comparison linking signal-adaptive decomposition, annotation-independent segmentation, traceable feature families, and recording-level control of segment leakage. The findings support EMD and fixed-window analysis as testable methodological choices for term-versus-preterm EHG classification rather than as a clinically validated risk model. Validation in larger external cohorts with explicit pregnancy linkage and defined prediction horizons is required before clinical interpretation.

\section*{Data availability}

The TPEHGT dataset used in this study is publicly available from PhysioNet, version 1.0.0, DOI: \url{https://doi.org/10.13026/C2166R}. The analysis code required to reproduce the reported analyses, including feature extraction, model training, cross-validation, recording-level aggregation, metric calculation, and figure generation, is available at \url{https://github.com/SenithJayakody/tpehgt-preterm-emd.git}.

\bibliography{plos_bibtex_sample}

@Article{Ohuma2023,
author={Ohuma, Eric O. and Moller, Ann-Beth and Bradley, Ellen and Chakwera, Samuel and Hussain-Alkhateeb, Laith and Lewin, Alexandra and Okwaraji, Yemisrach B. and Mahanani, Wahyu Retno and Johansson, Emily White and Lavin, Tina and Fernandez, Diana Estevez and Dom{\'i}nguez, Giovanna Gatica and de Costa, Ayesha and Cresswell, Jenny A. and Krasevec, Julia and Lawn, Joy E. and Blencowe, Hannah and Requejo, Jennifer and Moran, Allisyn C.},
title={National, regional, and global estimates of preterm birth in 2020, with trends from 2010: a systematic analysis},
journal={The Lancet},
year={2023},
volume={402},
number={10409},
pages={1261-1271},
issn={0140-6736},
doi={10.1016/S0140-6736(23)00878-4}
}

@misc{WHO_preterm_2023,
  author = {{World Health Organization}},
  title = {Preterm birth},
  year = {2023},
  note = {Accessed 7 May 2026},
  url = {https://www.who.int/news-room/fact-sheets/detail/preterm-birth}
}

@misc{WHO_newborn_2024,
  author       = {{World Health Organization}},
  title        = {Newborn mortality},
  year         = {2024},
  howpublished = {\url{https://www.who.int/news-room/fact-sheets/detail/newborn-mortality}},
  note         = {Accessed 30 June 2026}
}

@misc{UNICEF_neonatal_2026,
  author       = {{UNICEF}},
  title        = {Neonatal mortality},
  year         = {2026},
  month        = mar,
  day          = {1},
  howpublished = {\url{https://data.unicef.org/topic/child-survival/neonatal-mortality/}},
  note         = {Accessed 1 July 2026}
}

@misc{CDC_preterm_2024,
  author       = {{Centers for Disease Control and Prevention}},
  title        = {Preterm Birth},
  year         = {2024},
  howpublished = {\url{https://www.cdc.gov/maternal-infant-health/preterm-birth/index.html}},
  note         = {Accessed 30 June 2026}
}

@book{who2022tocolytic,
  author    = {{World Health Organization}},
  title     = {WHO Recommendation on Tocolytic Therapy for Improving Preterm Birth Outcomes},
  year      = {2022},
  publisher = {World Health Organization},
  address   = {Geneva},
  isbn      = {978-92-4-005722-7}
}

@article{ACOG2017_corticosteroids,
  author  = {{Committee on Obstetric Practice}},
  title   = {Committee Opinion No. 713: Antenatal Corticosteroid Therapy for Fetal Maturation},
  journal = {Obstetrics \& Gynecology},
  year    = {2017},
  volume  = {130},
  number  = {2},
  pages   = {e102--e109},
  doi     = {10.1097/AOG.0000000000002237}
}

@misc{NICE_NG25,
  author       = {{National Institute for Health and Care Excellence}},
  title        = {Preterm labour and birth},
  year         = {2015},
  howpublished = {\url{https://www.nice.org.uk/guidance/ng25}},
  note         = {NICE guideline NG25, last updated 10 June 2022; Accessed 30 June 2026}
}

@article{SonMiller2017,
  author  = {Son, Moeun and Miller, Emily S.},
  title   = {Predicting preterm birth: Cervical length and fetal fibronectin},
  journal = {Seminars in Perinatology},
  year    = {2017},
  volume  = {41},
  number  = {8},
  pages   = {445--451},
  doi     = {10.1053/j.semperi.2017.08.002}
}

@article{GarfieldManer2007,
  author = {Garfield, Robert E. and Maner, William L.},
  title = {Physiology and electrical activity of uterine contractions},
  journal = {Seminars in Cell and Developmental Biology},
  year = {2007},
  volume = {18},
  number = {3},
  pages = {289--295},
  doi = {10.1016/j.semcdb.2007.05.004}
}

@article{Lucovnik2011,
  author = {Lucovnik, Miha and Kuon, Ruben J. and Chambliss, Linda R. and Maner, William L. and Shi, Shao-Qing and Shi, Leili and Balducci, James and Garfield, Robert E.},
  title = {Use of uterine electromyography to diagnose term and preterm labor},
  journal = {Acta Obstetricia et Gynecologica Scandinavica},
  year = {2011},
  volume = {90},
  number = {2},
  pages = {150--157},
  doi = {10.1111/j.1600-0412.2010.01031.x}
}

@article{Buhimschi1997,
  author = {Buhimschi, Catalin and Boyle, Mary B. and Garfield, Robert E.},
  title = {Electrical activity of the human uterus during pregnancy as recorded from the abdominal surface},
  journal = {Obstetrics \& Gynecology},
  year = {1997},
  volume = {90},
  number = {1},
  pages = {102--111},
  doi = {10.1016/S0029-7844(97)83837-9}
}

@article{GarciaCasado2018_EHGReview,
  author  = {Garcia-Casado, J. and Ye-Lin, Y. and Prats-Boluda, G. and Mas-Cabo, J. and Alberola-Rubio, J. and Perales, A.},
  title   = {Electrohysterography in the diagnosis of preterm birth: a review},
  journal = {Physiological Measurement},
  year    = {2018},
  volume  = {39},
  number  = {2},
  doi     = {10.1088/1361-6579/aaad56}
}

@article{GarfieldSims1977,
  author = {Garfield, Robert E. and Sims, Stephen and Daniel, Edwin E.},
  title = {Gap junctions: their presence and necessity in myometrium during parturition},
  journal = {Science},
  year = {1977},
  volume = {198},
  number = {4320},
  pages = {958--960},
  doi = {10.1126/science.929182}
}

@article{Aguilar2010,
  author = {Aguilar, Hector N. and Mitchell, Bryan F.},
  title = {Physiological pathways and molecular mechanisms regulating uterine contractility},
  journal = {Human Reproduction Update},
  year = {2010},
  volume = {16},
  number = {6},
  pages = {725--744},
  doi = {10.1093/humupd/dmq016}
}

@article{garfield1998control,
  author    = {Garfield, Robert E. and Saade, George and Buhimschi, Catalin and 
               Buhimschi, Irina and Shi, Li and Shi, Song Q. and Chwalisz, Klaus},
  title     = {Control and Assessment of the Uterus and Cervix During Pregnancy and Labour},
  journal   = {Human Reproduction Update},
  year      = {1998},
  volume    = {4},
  number    = {5},
  pages      = {673--695},
  month     = {Sep-Oct},
  doi       = {10.1093/humupd/4.5.673},
  pmid      = {10027621}
}

@incollection{garfield1993control,
  author    = {Garfield, Robert E. and Yallampalli, Chandrasekhar},
  title     = {Control of Myometrial Contractility and Labour},
  booktitle = {Basic Mechanisms Controlling Term and Preterm Labour},
  editor    = {Chwalisz, Klaus and Garfield, Robert E.},
  series    = {Ernst Schering Research Foundation Workshop},
  volume    = {7},
  publisher = {Springer-Verlag},
  address   = {Berlin, Heidelberg, New York},
  year      = {1993},
  pages     = {1--29}
}

@article{wolfs1979electromyographic,
  author    = {Wolfs, G. M. and van Leeuwen, M.},
  title     = {Electromyographic Observations on the Human Uterus During Labour},
  journal   = {Acta Obstetricia et Gynecologica Scandinavica},
  year      = {1979},
  volume    = {90},
  number    = {suppl.},
  pages     = {1--61}
}

@incollection{csapo1981force,
  author    = {Csapo, A. I.},
  title     = {Force of Labour},
  booktitle = {Principles and Practice of Obstetrics and Perinatology},
  editor    = {Iffy, L. and Kaminetzky, H. A.},
  volume    = {2},
  publisher = {John Wiley \& Sons},
  address   = {New York},
  year      = {1981},
  pages     = {761--799}
}

@article{Devedeux1993,
  author  = {Devedeux, D. and Marque, C. and Mansour, S. and Germain, G. and Duchene, J.},
  title   = {Uterine electromyography: A critical review},
  journal = {American Journal of Obstetrics and Gynecology},
  year    = {1993},
  volume  = {169},
  number  = {6},
  pages   = {1636--1653},
  doi     = {10.1016/0002-9378(93)90456-S}
}

@article{li2022uterine,
  author    = {Li, P. and Huang, Q. and Wang, L. and Garfield, Robert E. and Liu, H.},
  title     = {Uterine Electrical Signals and Cervical Dilation During the First Stage of Labor},
  journal   = {Archives of Obstetrics and Gynecology},
  year      = {2022},
  volume    = {3},
  number    = {1},
  pages     = {47--52},
  doi = {10.33696/Gynaecology.3.028}
}

@article{mascabo2019uterine,
  author  = {Mas-Cabo, Javier and Prats-Boluda, Gema and Perales, Alfredo and Garcia-Casado, Javier and Alberola-Rubio, Jos{\'e} and Ye-Lin, Yiyao},
  title   = {Uterine Electromyography for Discrimination of Labor Imminence in Women with Threatened Preterm Labor under Tocolytic Treatment},
  journal = {Medical \& Biological Engineering \& Computing},
  year    = {2019},
  volume  = {57},
  number  = {2},
  pages   = {401--411},
  doi     = {10.1007/s11517-018-1888-y}
}

@article{FeleZorz2008,
  author = {Fele-{\v{Z}}or{\v{z}}, Ga{\v{s}}per and Kav{\v{s}}ek, Gorazd and Novak-Antoli{\v{c}}, {\v{Z}}iva and Jager, Franc},
  title = {A comparison of various linear and non-linear signal processing techniques to separate uterine EMG records of term and pre-term delivery groups},
  journal = {Medical \& Biological Engineering \& Computing},
  year = {2008},
  volume = {46},
  number = {9},
  pages = {911--922},
  doi = {10.1007/s11517-008-0350-y}
}

@article{Jager2018_TPEHGT,
  author = {Jager, Franc and Liben{\v{s}}ek, Sonja and Ger{\v{s}}ak, Ksenija},
  title = {Characterization and automatic classification of preterm and term uterine records},
  journal = {PLOS ONE},
  year = {2018},
  volume = {13},
  number = {8},
  pages = {e0202125},
  doi = {10.1371/journal.pone.0202125}
}

@article{Fergus2013_EHGML,
  author  = {Fergus, Paul and Cheung, P. and Hussain, Abir J. and Al-Jumeily, Dhiya and Dobbins, Chelsea and Iram, Sarah},
  title   = {Prediction of Preterm Deliveries from EHG Signals Using Machine Learning},
  journal = {PLOS ONE},
  year    = {2013},
  volume  = {8},
  number  = {10},
  pages   = {e77154},
  doi     = {10.1371/journal.pone.0077154}
}

@article{Alamedine2013_EHGFeatureSelection,
  author  = {Alamedine, David and Khalil, Mohamad and Marque, Catherine},
  title   = {Comparison of Different EHG Feature Selection Methods for the Detection of Preterm Labor},
  journal = {Computational and Mathematical Methods in Medicine},
  year    = {2013},
  volume  = {2013},
  pages   = {485684},
  doi     = {10.1155/2013/485684}
}

@article{RomeroMorales2023,
  author  = {Romero-Morales, H{\'e}ctor and Mu{\~n}oz-Montes de Oca, Jenny Noem{\'i} and Mora-Mart{\'i}nez, Rodrigo and Mina-Paz, Yecid and Reyes-Lagos, Jos{\'e} Javier},
  title   = {Enhancing classification of preterm-term birth using continuous wavelet transform and entropy-based methods of electrohysterogram signals},
  journal = {Frontiers in Endocrinology},
  year    = {2023},
  volume  = {13},
  pages   = {1035615},
  doi     = {10.3389/fendo.2022.1035615}
}

@article{Goldsztejn2023_EHGDeepLearning,
  author  = {Goldsztejn, Uri and Nehorai, Arye},
  title   = {Predicting preterm births from electrohysterogram recordings via deep learning},
  journal = {PLOS ONE},
  year    = {2023},
  volume  = {18},
  number  = {5},
  pages   = {e0285219},
  doi     = {10.1371/journal.pone.0285219}
}

@article{Fischer2023_EHGDeepLearning,
  author  = {Fischer, A. M. and Vullings, R. and Gommers, J. S. M. and Oei, S. G. and Bergmans, J. W. M. and Mischi, M.},
  title   = {End-to-end learning with interpretation on electrohysterography data to predict preterm birth},
  journal = {Computers in Biology and Medicine},
  year    = {2023},
  volume  = {158},
  pages   = {106846},
  doi     = {10.1016/j.compbiomed.2023.106846}
}

@article{Pirnar2024,
  author = {Pirnar, Ziga and Jager, Franc and Gersak, Ksenija},
  title = {Peak amplitude of the normalized power spectrum of the electromyogram of the uterus in the low frequency band is an effective predictor of premature birth},
  journal = {PLOS ONE},
  year = {2024},
  volume = {19},
  number = {9},
  pages = {e0308797},
  doi = {10.1371/journal.pone.0308797}
}

@article{Vandewiele2021,
  author = {Vandewiele, Gilles and Dehaene, Isabelle and Kovacs, Gyorgy and Sterckx, Lucas and Janssens, Olivier and Ongenae, Femke and De Backere, Femke and De Turck, Filip and Roelens, Kristien and Decruyenaere, Johan and Van Hoecke, Sofie and Demeester, Thomas},
  title = {Overly optimistic prediction results on imbalanced data: A case study of flaws and benefits when applying over-sampling},
  journal = {Artificial Intelligence in Medicine},
  year = {2021},
  volume = {111},
  pages = {101987},
  doi = {10.1016/j.artmed.2020.101987}
}

@article{Huang1998_EMD,
  author = {Huang, Norden E. and Shen, Zheng and Long, Steven R. and Wu, Manli C. and Shih, Hsing H. and Zheng, Quanan and Yen, Nai-Chyuan and Tung, Chi Chao and Liu, Henry H.},
  title = {The empirical mode decomposition and the Hilbert spectrum for nonlinear and non-stationary time series analysis},
  journal = {Proceedings of the Royal Society of London. Series A: Mathematical, Physical and Engineering Sciences},
  year = {1998},
  volume = {454},
  number = {1971},
  pages = {903--995},
  doi = {10.1098/rspa.1998.0193}
}

@article{Pollard2026PhysioNet,
  author  = {Pollard, Tom and Moody, Benjamin E. and Lehman, Li-wei H. and Gow, Brian J. and Fernandes, Chrystinne and Xie, Chen and Johnson, Alistair and Mark, Roger G. and Heldt, Thomas},
  title   = {PhysioNet as a global platform for biomedical research},
  journal = {Nature Health},
  year    = {2026},
  volume  = {1},
  pages   = {792--795},
  doi     = {10.1038/s44360-026-00096-z}
}

@ARTICLE{Welch1967,
  author={Welch, P.},
  journal={IEEE Transactions on Audio and Electroacoustics}, 
  title={The use of fast Fourier transform for the estimation of power spectra: A method based on time averaging over short, modified periodograms}, 
  year={1967},
  volume={15},
  number={2},
  pages={70-73},
  doi={10.1109/TAU.1967.1161901}
}

@article{Lammers2013,
  author = {Lammers, Wim J. E. P.},
  title = {The electrical activities of the uterus during pregnancy},
  journal = {Reproductive Sciences},
  year = {2013},
  volume = {20},
  number = {2},
  pages = {182--189},
  doi = {10.1177/1933719112446082}
}

@article{janjarasjitt2021preterm,
  author    = {Janjarasjitt, S.},
  title     = {Preterm-Term Birth Classification Using EMD-Based Time-Domain Features of Single-Channel Electrohysterogram Data},
  journal   = {Physical and Engineering Sciences in Medicine},
  year      = {2021},
  volume    = {44},
  number    = {4},
  pages     = {1151--1159},
  month     = {Dec},
  doi       = {10.1007/s13246-021-01051-w},
  pmid      = {34463948}
}

@article{janjarasjitt2022comparison,
  author    = {Janjarasjitt, S.},
  title     = {Comparison of Wavelet-Based Decomposition and Empirical Mode Decomposition of Electrohysterogram Signals for Preterm Birth Classification},
  journal   = {ETRI Journal},
  year      = {2022},
  volume    = {44},
  number    = {5},
  pages     = {826--836},
  doi       = {10.4218/etrij.2021-0220}
}

@article{mohammadifar2022prediction,
  author    = {Mohammadi Far, S. and Beiramvand, M. and Shahbakhti, M. and Augustyniak, P.},
  title     = {Prediction of Preterm Delivery from Unbalanced EHG Database},
  journal   = {Sensors},
  year      = {2022},
  volume    = {22},
  number    = {4},
  pages     = {1507},
  doi       = {10.3390/s22041507}
}

@article{mohammadifar2023prediction,
  author    = {Mohammadi Far, S. and Beiramvand, M. and Shahbakhti, M. and Augustyniak, P.},
  title     = {Prediction of Preterm Labor from the Electrohysterogram Signals Based on Different Gestational Weeks},
  journal   = {Sensors},
  year      = {2023},
  volume    = {23},
  number    = {13},
  pages     = {5965},
  doi       = {10.3390/s23135965}
}

@inproceedings{Kaiser1990,
  author = {Kaiser, James F.},
  title = {On a simple algorithm to calculate the energy of a signal},
  booktitle = {International Conference on Acoustics, Speech, and Signal Processing},
  year = {1990},
  pages = {381--384},
  doi = {10.1109/ICASSP.1990.115702}
}

@article{BandtPompe2002,
  author = {Bandt, Christoph and Pompe, Bernd},
  title = {Permutation entropy: A natural complexity measure for time series},
  journal = {Physical Review Letters},
  year = {2002},
  volume = {88},
  number = {17},
  pages = {174102},
  doi = {10.1103/PhysRevLett.88.174102}
}

@article{RichmanMoorman2000,
  author = {Richman, Joshua S. and Moorman, J. Randall},
  title = {Physiological time-series analysis using approximate entropy and sample entropy},
  journal = {American Journal of Physiology-Heart and Circulatory Physiology},
  year = {2000},
  volume = {278},
  number = {6},
  pages = {H2039--H2049},
  doi = {10.1152/ajpheart.2000.278.6.H2039}
}

@inproceedings{Prokhorenkova2018CatBoost,
  author    = {Prokhorenkova, Liudmila and Gusev, Gleb and Vorobev, Aleksandr and Dorogush, Anna Veronika and Gulin, Andrey},
  title     = {CatBoost: unbiased boosting with categorical features},
  booktitle = {Advances in Neural Information Processing Systems},
  year      = {2018},
  volume    = {31},
  pages     = {6638--6648}
}

@article{Chicco2020_MCC,
  author  = {Chicco, Davide and Jurman, Giuseppe},
  title   = {The advantages of the Matthews correlation coefficient ({MCC}) over {F1} score and accuracy in binary classification evaluation},
  journal = {BMC Genomics},
  year    = {2020},
  volume  = {21},
  number  = {1},
  pages   = {6},
  doi     = {10.1186/s12864-019-6413-7}
}

@article{Ren2015EMD,
  author = {Ren, Peng and Yao, Sa and Li, Jun and Valdes-Sosa, Pedro A. and Kendrick, Keith M.},
  title = {Improved prediction of preterm delivery using empirical mode decomposition analysis of uterine electromyography signals},
  journal = {PLOS ONE},
  year = {2015},
  volume = {10},
  number = {7},
  pages = {e0132116},
  doi = {10.1371/journal.pone.0132116}
}

@article{Cui2025MEMD,
  author  = {Cui, Jiawen and Zhang, Xu and Li, Xinhui and Luo, Xuanyu and Chen, Xiang and Yin, Zongzhi},
  title   = {Preterm birth prediction from electrohysterogram using multivariate empirical mode decomposition},
  journal = {Medical \& Biological Engineering \& Computing},
  year    = {2025},
  volume  = {63},
  number  = {6},
  pages   = {1867--1880},
  month   = jun,
  doi     = {10.1007/s11517-025-03293-2},
  pmid    = {39893327}
}

\end{document}